\documentclass{aa}

\usepackage{amsmath}
\usepackage{amssymb}
\usepackage{graphicx}

\usepackage{color}
\usepackage{natbib}
\usepackage{graphicx}
\usepackage{txfonts}
\usepackage[]{latexsym}
\usepackage[english]{babel}
\usepackage{graphicx}
\usepackage{hyperref}
\usepackage{color}
\usepackage{bm}

\usepackage{tikz}

\usepackage{soul}

\def\apjl{ApJL}
\def\apjs{ApJS}

\def\aap{{A\&A}}

\newcommand{\T}{\hat{\mathcal{T}}}

\newcommand{\Dk}[1]{\frac{d^3#1}{(2\pi)^3}}

\newcommand{\ve}[1]{{\text{\bf #1}}} 
\newcommand{\vk}{\ve k}
\newcommand{\vp}{\ve p}
\newcommand{\vq}{\ve q}
\newcommand{\vx}{\ve x}

\newcommand{\Ps}{\mathbf{\Psi}}

\newcommand{\obd}{\omega_\text{BD}}

\newcommand{\ikk}{\underset{\vk_{12}= \vk}{\int}}
\newcommand{\ikkk}{\underset{\vk_{123}= \vk}{\int}}

\newcommand{\bvarphi}{\bar{\varphi}}

\newcommand{\Mp}{M_\text{Pl}}
\begin{document} 

   \title{Screenings in Modified Gravity: a perturbative approach}
  
   \author{Alejandro Aviles
          \inst{1,2} \and Jorge L. Cervantes-Cota \inst{2}
          \and
          David F. Mota \inst{3}
          }
   \institute{Consejo Nacional de Ciencia y Tecnolog\'ia, Av. Insurgentes Sur 1582,
Colonia Cr\'edito Constructor, Del. Benito Jurez, 03940, Ciudad de M\'exico, M\'exico  \and Departamento de F\'isica, Instituto Nacional de Investigaciones Nucleares,
Apartado Postal 18-1027, Col. Escand\'on, Ciudad de M\'exico,11801, M\'exico. \and Institute Of Theoretical Astrophysics, University of Oslo, 0315 Oslo, Norway}

   \date{Received...; accepted...}

  \abstract{
   {}
   
   We present a formalism to study screening mechanisms in modified theories of gravity via perturbative methods in different cosmological scenarios. 
   We consider Einstein frame posed theories that are recast as Jordan frame theories, where a known formalism is employed, 
    though the resulting non-linearities of the Klein-Gordon equation acquire an explicit coupling between matter and the scalar field, 
    which is not present in Jordan frame theories.
    The obtained growth functions are then separated in screening and non-screened contributions to facilitate its analysis. 
    This allows us to compare several theoretical models and to recognize patterns which can be used to differentiate models and 
    their screening mechanisms. In particular, we find anti-screening features in the Symmetron model.   In opposition, chameleon type theories, 
    both in the Jordan and in the Einstein frame, always present a screening behaviour. Up to third order in perturbation, we 
    find no anti-screening behaviour in theories with a Vainshtein mechanism, such as the DGP and the cubic Galileon.

   {}
   {}
   {}
   {}}


   \maketitle
%

\section{Introduction}

Two decades have passed since the discovery of the present days acceleration of the Universe \citep{Riess1998AJ....116.1009R,Perlmutter1999ApJ...517..565P}. Its physical origin, however, is still a mystery.  The
simplest choice is that such acceleration is the result of the vacuum energy, in the form of a cosmological constant, in Einstein's equations. 
This model is in agreement with current observations, but  is plagued by the fine-tuning problem. Another simple option is  that dark energy is 
a dynamical field, as in quintessence models. This field may also have interactions with the  dark matter  sector giving rise  to interacting  dark energy models \citep{Amendola2010deto.book.....A}.

Instead of modifying the particle content of the Universe, an alternative solution is to extend Einstein gravity in such a way that the present 
acceleration of the Universe is accounted for. However, precise measurements on Earth, in the solar system, in binary pulsars, and of gravitational 
waves \citep{2016arXiv160203841T,Berotti2003Natur.425..374B, Will2006LRR.....9....3W, Williams2004PhRvL..93z1101W}, strongly constrain deviations 
from general relativity (GR) at those scales \citep{2014LRR....17....4W}. This is a challenge to theories of gravity beyond Einstein GR, that claim 
to be an explanation of the present days acceleration \citep{clifton_modified_2012, 2015PhR...568....1J, 2015arXiv151205356B}. 

Notice however that all these tests are probing extensions to GR in astrophysical systems that reside in dense galactic environments. Conditions are 
therefore far from the cosmological background density, curvature, and even the gravitational potential differs from the background one by several orders of magnitude \citep{2015ApJ...802...63B}. 

A way of evading the high density environmental constraints, while still allowing deviations from GR on cosmological scales, is to hide modifications 
to Einstein's gravity in those environments. The idea is that one modifies the geometry by introducing a new degree of freedom which drives the Universe 
acceleration. Such degree of freedom, in the simplest case a scalar field, would however be suppressed in high density/curvature environments.    

There are several screening mechanisms proposed in the literature, and there are several ways of classifying them \citep{Joyce:2014kja,brax_screening_2013,brax_distinguishing_2015,Koyama:2015vza}. 
In this work we investigate screening mechanisms which result from one of these three properties: i) {\it Weak coupling}, in which the coupling to matter fields is small in regions of high density, 
hence suppressing the fifth force. At large scales the density is small and the fifth forces acts. Examples  of theories of this type are Symmetron \citep{PhysRevLett.104.231301,PhysRevD.72.043535,PhysRevD.77.043524}
and varying  dilaton \citep{DAMOUR1994532,PhysRevD.83.104026}; ii)  {\it Large mass}, when mass of the fluctuation is large in regions of high density, suppressing the fifth force, 
but at low densities the scalar field is small and can mediate a fifth force.  Examples of this type are Chameleons \citep{Khoury:2003aq,khoury:2003rn}; and iii) {\it Large inertia}, 
when the scalar field kinetic function depends on the environment, making it large in density regions, and then the coupling to matter is suppressed. There are two cases, when either 
first or second derivatives of the scalar field are large. Examples of the former are $K$-mouflage models \citep{Babichev:2009ee,PhysRevD.84.061502}, and of the latter are Vainshtein models \citep{Vainshtein:1972sx}.

Screening mechanisms are a relatively generic prediction of viable modified gravity (MG) theories \citep{2012PhRvD..86d4015B}. Therefore detecting them would be a signature 
of beyond GR physics. Observational tests of screening focus on the transition between the fully screened and unscreened regimes, where deviations from GR are expected to be 
most pronounced. For viable cosmological models, such transition occurs in regions where the matter density and gravitational potential are nonlinear and start to approach 
their linear or background values: this can be observed in the outskirts of dark matter halos and its properties 
\citep[e.g.][]{2005PhRvD..71f4030S,davis, 2008PhRvD..78l3524O, 2009PhRvD..79h3518S, 2013JCAP...10..027B, 2013PhRvD..88h4029W, 2014ApJS..211...23Z, clifton,2015PhRvL.114y1101L, 2009arXiv0911.1829M, 2012PhRvD..85j2001L, 2013MNRAS.431..749C, 2013PhRvL.110o1104L, 2013JCAP...10..012H, gronke_gravitational_2014, gronke_halos_2015, 2016arXiv160300056S}.

Screening mechanisms are in fact a nonlinear effect. Therefore, predictions of their signatures in astrophysical systems are 
computed integrating the fully nonlinear equations of motion for the gravity extra-degree of freedom and using also N-body 
simulations in order to simulate nonlinear structure formation. These computations are however extremely expensive time wise. 
It is therefore not viable to use them to do parameter estimation or even putting constraints on the parameter space of such 
theories using nonlinear codes of structure formation. It is then imperative to find other methods to describe and to probe screening mechanisms in faster and accurate ways.
Although screenings are more usually studied in the highly non-linear regime, they leave imprints in quasi-linear scales that can be captured by 
cosmological perturbation theory (PT) and should be considered in the low density regions of large scale structure formation; see for example \citep{2009PhRvD..79l3512K}.

On the other hand, PT has experienced many developments in recent years \citep{Matsubara:2007wj,Baumann:2010tm,Carlson:2012bu} in part because 
it can be useful to analytically understand different effects in the power spectrum and correlation function for the dark matter clustering.
These effects can be confirmed or not, and further explored, with simulations to ultimately understand the outcomes of present and future galaxy surveys, 
such as eBOSS \citep{Zhao:2015gua}, DESI \citep{Aghamousa:2016zmz}, EUCLID \citep{Amendola2013}, LSST \citep{Abate:2012za}, among others. 
There are mainly two approaches to study PT: Eulerian standard PT (SPT) and Lagrangian PT (LPT), both with advantages and drawbacks, but at 
the end they are complementary \citep{Tassev:2013rta}. The nonlinear PT for MG was developed initially in \citep{2009PhRvD..79l3512K}, and further studied in several other works 
\citep{Taruya:2013quf,Brax:2013fna,Taruya:2014faa,Bellini:2015oua,Taruya:2016jdt,Bose:2016qun,Barrow:2002zh,Akrami:2013ffa,Fasiello:2017bot,Bose:2017dtl,Aviles:2017aor,Hirano:2018uar,Bose:2018orj,Bose:2018zpk,Aviles:2018saf}.  The LPT for dark matter fluctuations in MG was developed in \citep{Aviles:2017aor}, and further studies for biased tracers in \citep{Aviles:2018saf}.   Having PT for MG at hand has the advantage to allow us to understand the role of that physical parameters play in the screening features of dark matter statistics. In the present work, we aim at studying some of these effects through screening mechanisms by studying them at second and third order  perturbation levels using PT for some MG models. 
To this end we build on the LPT formalism developed in \cite{Aviles:2017aor}, initially posed for MG theories in the Jordan frame, 
in order to apply it to theories in the Einstein frame. 
Due to a direct coupling of the scalar field and the dark matter in the Klein-Gordon equation, the equations that govern the screening 
can differ substantially than those in Jordan frame MG theories. 
In general screening effects depend on the type of nonlinearities introduced in the Lagrangian density. We present a detailed analysis 
of screening features and identify the theoretical roots of its origin. Our results show that screenings possess peculiar features that 
depend on scalar field effective mass and couplings, and that may in particular cases drive to anti-screening effects in the power spectrum, 
as e.g. in the Symmetron. We perform this analysis by separating the growth functions in screening and non-screened pieces. 
Notice however, that in this paper we do not compare the perturbative approach with a fully nonlinear simulation. We refer the reader 
to see for instance \citep{2009PhRvD..79l3512K} at such investigations at the level of the power spectrum.

This work is organized as follows: in section \ref{sec:PT} we set up the formalism to do perturbation theory in both the Einstein and Jordan 
frames; in section \ref{sec:MGmodels} we apply such methods to the specific gravity models investigated here; in section \ref{sec:PS} we show 
the matter power spectra and section \ref{sec:Growth} the screening growth functions analysis. We conclude in section \ref{sec:conclusions} 
with a discussion of our results. Some formulae are displayed in appendix \ref{app:3rdOrder}.

\begin{section}{Perturbation theory in the Einstein frame}\label{sec:PT}


In this section we are interested in MG theories defined in the Einstein frame with action
\begin{equation}\label{EF_action}
 S = \int d^4x \sqrt{-g} \left( \frac{\Mp^2}{2} R - (\nabla \varphi)^2 - V(\varphi) \right)  + S_m[\tilde{g}_{\mu\nu}],
\end{equation}
with the conformal metric
\begin{equation}
 \tilde{g}_{\mu\nu}=(C(\varphi))^2g_{\mu\nu},
\end{equation}
where $C(\varphi)$ is a conformal factor (in the literature is more common to find $A(\varphi)$;
we use $C$ instead because $A$ will be used below to characterize the 
strength of the gravitational force). 
By taking variations of the action (\ref{EF_action}) with respect to the scalar field we obtain the Klein-Gordon equation
\begin{equation}\label{KGeq}
 \square \varphi = \frac{d V_{eff}}{d \varphi},  \qquad V_{eff} = V(\varphi) -  T ( C(\varphi) - 1),
\end{equation}
where $T$ is the trace of the energy momentum tensor of matter.
In PT we split the scalar field in background $\bvarphi$ and perturbed $\delta \varphi$ pieces
\begin{equation}
 \varphi(\vx,t) = \bar{\varphi}(t) + \delta \varphi(\vx,t). 
\end{equation}
Hereafter a bar over a dynamical quantity means we are referring to its homogeneous and isotropic, background value; we also assume a dark matter perfect fluid with $T = - \, \rho$.  
In the following we will adopt the quasi-static limit for the perturbed piece which relies on 
neglecting temporal derivatives in the Klein-Gordon equation, thus Eq.~(\ref{KGeq}) becomes
\begin{align}\label{KGeqPert}
\frac{1}{a^2} \nabla^2_\vx \delta \varphi 
 &= \sum_{n=1}^{\infty} \frac{1}{n!}\left[V^{(n+1)}(\bvarphi) + \bar{\rho} C^{(n+1)} (\bvarphi) \right] (\delta \varphi)^n + \nonumber\\
& + \bar{\rho} \delta \sum_{n=0}^{\infty} \frac{1}{n!}  C^{(n+1)} (\bvarphi)  (\delta \varphi)^n  \nonumber\\
 &=  \frac{\beta}{\Mp} \bar{\rho} \delta + m^2(\bvarphi) \delta \varphi + \sum_{n=2}^{\infty} \frac{1}{n!} \Mp^{1-n} \kappa_{n+1} (\delta \varphi)^n + \nonumber\\
    &+ \bar{\rho} \delta \sum_{n=1}^{\infty} \frac{1}{n!} \Mp^{-1-n} \beta_{n+1} (\delta \varphi)^n,
\end{align}
where we have subtracted the background evolution.
Here $V^{(n)}(\bvarphi)$ and $C^{(n)}(\bvarphi)$ denote the \emph{n-\'esime} derivative of $V$ and $C$ functions evaluated at background values. 
In the above equation we also introduced the matter overdensity
$\delta$, defined through $\rho(\vx,t) = \bar{\rho}(t) (1 + \delta(\vx,t))$. 
We also have introduced, following  (\cite{Brax:2013fna}),
\begin{equation}
 \kappa_n(\bvarphi) = \Mp^{n-2} ( V^{(n)}(\bvarphi) + \bar{\rho} C^{(n)}(\bvarphi) ), \qquad m^2 = \kappa_2.
\end{equation}
\begin{equation}
\beta_n(\bvarphi) = \Mp^n C^{(n)}(\bvarphi), \qquad \beta = \beta_1.
\end{equation}
We will work in Lagrangian space, where the position $\vx$ of a dark matter particle, or fluid element, with initial Lagrangian coordinate $\vq$, is given by
\begin{equation} \label{LtoEpos}
 \vx(\vq,t) = \vq + \Ps(\vq,t),
\end{equation}
where $\Ps$ is the Lagrangian displacement vector field. We further assume that $\Ps$ is longitudinal 
and that it is a Gaussian distributed variable at linear order. 
Dark matter particles follow geodesics of the conformal metric $\tilde{{\mathbf g}}$,
\begin{equation} \label{GE}
 \ddot{\Ps} + 2 H \dot{\Ps}   = -\frac{1}{a^2}\nabla_\vx \psi_\text{N} - \frac{1}{a^2}\nabla_\vx \log C(\varphi) ,
\end{equation}
where $\psi_\text{N}$ denotes the Newtonian potential which obeys the Poisson equation\footnote{We use the 
notation $\nabla_{\vx} = \partial/\partial \vx$ for derivatives with respect to Eulerian coordinates. 
For spatial  Lagrangian coordinates derivatives we use $\nabla =  \partial/\partial \vq$.}
\begin{equation} \label{PE}
 \nabla^2_\vx \psi_\text{N} = 4 \pi G a^2 \bar{\rho} \delta(\vx).
\end{equation}
A noticeable difference between MG theories defined in the Einstein and Jordan frames is that in the latter case the new, scalar degree of freedom sources the 
Poisson equation instead of the geodesic equation.

Equation (\ref{LtoEpos}) can be regarded as a coordinate transformation between Lagrangian and Eulerian coordinates, 
with Jacobian matrix $J_{ij}= (\partial x^i/ \partial q^j) = \delta_{ij} + \Psi_{i,j}$ and Jacobian determinant
$J= \text{det}(J_{ij})$. From mass conservation a relation between matter overdensities and Lagrangian displacement can be obtained (\cite{BouColHivJus95})
\begin{equation}\label{Jtodelta}
 \delta(\vx,t) = \frac{1-J(\vq,t)}{J(\vq,t)}.
\end{equation}
The set of equations (\ref{KGeq}), (\ref{GE}) and (\ref{PE}) will be treated perturbatively in order to solve for the displacement field.
But instead of working with the field $\delta \varphi$, we find convenient to define the rescaled field
\begin{equation} \label{Defchi}
 \chi(\vq,t) \equiv -\frac{2 \beta}{C} \frac{\delta \varphi(\vq,t)}{\Mp},
\end{equation}
hereafter we denote $C \equiv C(\bvarphi)$ unless otherwise is explicitly stated. Note that Eq.~(\ref{Defchi}) is not a conformal transformation since the functions $\beta$ and $C$ are not free functions but they are evaluated at the background.
By taking the divergence of Eq.~(\ref{GE}) we have
\begin{equation} \label{pre_em}
 \nabla_\vx \cdot \Big( \ddot{\Ps} + 2 H \dot{\Ps} \Big)  = -4 \pi G \bar{\rho} \delta(\vx) + \frac{1}{2a^2}\nabla^2_\vx \chi, 
\end{equation}
that has the structure of a MG theory in the Jordan frame, and the PT formalism 
developed in \cite{Aviles:2017aor} applies directly. 
In Lagrangian space we work in $q$-Fourier space, in which the transformation is taken with respect to $q$-coordinates, thus 
when transforming gradients with respect to $x$-coordinates, \emph{frame-lagging} terms are introduced

\begin{equation}
 \nabla_\vx^2 \chi = \nabla^2 \chi  + (\nabla_\vx^2 \chi -\nabla^2 \chi ).
\end{equation}
These frame-lagging terms are necessary to obtain the correct limit of the theory at large scales, particularly for those theories in which the associated 
fifth force is short-ranged and the $\Lambda$CDM limit at large scales should be recovered. Since at linear order spatial derivatives with respect to 
Eulerian and Lagrangian coordinates coincide, the frame-lagging gives nonlinear contributions to the theory.
Now, to make contact with other works it is worthy to introduce the quantities:
\begin{align}
 A(k,a) &=  A_{0}\left(1 + \frac{2 \beta^2}{C}\frac{ k^2/a^2}{k^2/a^2 + m^2}\right), \\
 A_{0}(a) &=  4\pi G \bar{\rho}, \\
 \Pi(k,a) 
 &= \frac{C}{6 a^2 \beta^2}\left(  k^2 +  m^2 a^2  \right), \\
 3 &+ 2 \obd(a) = \frac{C}{2\beta^2}, \\
 M_1(a) &= \frac{C}{2\beta^2} m^2.
\end{align}
The above equations can be used as a translation table in between different PT works in MG. Particularly in \cite{Aviles:2017aor}, the $M_1$ and $\obd$ functions, instead of $\beta$ and $m$, are used extensively. 

Now, let us come back to the Klein-Gordon equation, which  in $q$-Fourier space, for $\chi$ field is 
\begin{align}\label{KGeq_chi}
- \frac{k^2}{2 a^2} \chi(\vk) &= -(A(k)-A_0) \tilde{\delta}(\vk) + \frac{k^2/a^2}{6 \Pi(k)} \delta \mathcal{I}(\chi) - \nonumber\\
&- \frac{C}{2 \beta^2}\frac{k^2/a^2}{3 \Pi(k)} \frac{1}{2a^2} [(\nabla^2_\vx \chi - \nabla^2 \chi)](\vk),
\end{align}
where $[(\cdots)](\vk)$ means $q$-Fourier transform of $(\cdots)(\vq)$, and we also make notice that 
 $\tilde{\delta}(\vk) \equiv \int d^3 q e^{-i\vk \cdot \vq} \delta(\vx)$. 
To avoid confusion with the $q$-Fourier transform of $\delta(\vq)$ or the $x$-Fourier transform of 
$\delta(\vx)$, we write a tilde over that overdensity. Equation (\ref{KGeq_chi}) is derived directly using 
Eq.~(\ref{Defchi}), the second term in the right hand side (RHS) encodes all nonlinear terms of Eq.~(\ref{KGeqPert}), while the last term arises
when transforming derivatives from Eulerian to Lagrangian coordinates. Specifically, the contribution from screenings is given by
\begin{align}\label{dI}
\delta \mathcal{I}(\vk) &= \sum_{n=2}^\infty \frac{(-1)^{n+1}}{ 2^n n!} \frac{C^n \kappa_{n+1}}{\beta^{n+1}} [ \chi^n](\vk)+\nonumber \\
& + \sum_{n=1}^\infty \frac{(-1)^{n+1}}{2^n n! } \frac{2 A_0 C^n \beta_{n+1}}{\beta^{n+1}}[\chi^n \delta  ](\vk).
\end{align}

We formally expand quantities as\footnote{In this work we adopt the shorthand notations
\begin{equation} \label{int_not}
 \underset{\vk_{1\cdots n}= \vk}{\int} = \int \Dk{\vk_1} \cdots \Dk{\vk_n} (2 \pi)^3 \delta_\text{D}(\vk - \vk_{1\cdots n}),
\end{equation}
and $
 \vk_{1\cdots n} = \vk_1 + \cdots 
 +\vk_n$.
}
\begin{eqnarray}
\label{dIexp}
\delta \mathcal{I}(\vk) &=& \frac{1}{2} \ikk M_2(\vk_1,\vk_2) \chi(\vk_1)\chi(\vk_2)+\nonumber \\
    &+& \frac{1}{6} \ikkk M_3(\vk_1,\vk_2,\vk_3) \chi(\vk_1)\chi(\vk_2)\chi(\vk_3) + \cdots,
\end{eqnarray}
\begin{eqnarray}
&-\frac{1}{2a^2} [(\nabla^2_\vx \chi - \nabla^2 \chi)](\vk)= \frac{1}{2} \ikk \mathcal{K}^{(2)}_\text{FL}(\vk_1,\vk_2) \delta_L(\vk_1)\delta_L(\vk_2)+\nonumber \\
    &+ \frac{1}{6} \ikkk \mathcal{K}^{(3)}_\text{FL}(\vk_1,\vk_2,\vk_3)  \delta_L(\vk_1)\delta_L(\vk_2) \delta_L(\vk_3) + \cdots.
\end{eqnarray}
Expressions for the frame lagging kernels are derived in \cite{Aviles:2017aor}, for example to second order we have
\begin{align} \label{K2FL}
 \mathcal{K}^{(2)}_\text{FL}(\vk_1, \vk_2) &= 2 x^2 (A(k_1) +A(k_2) - 2 A_0) +  
  x\frac{k_2}{k_1} (A(k_1)  -  A_0) \nonumber \\ &+ x\frac{k_1}{k_2} (A(k_2)  -  A_0),
\end{align}
with $x=\hat{\vk}_1 \cdot \hat{\vk}_2$.

There is a crucial difference between theories in the Jordan and Einstein frames. In the latter there is a direct coupling 
$\chi \delta$ between the scalar field and the matter density, 
as can be seen from Eq.~(\ref{KGeq}) or from Eq.~(\ref{dI}). This
leads us to expand the overdensity in terms of the scalar field as
\begin{equation}\label{dchiexp}
\tilde{\delta}(\vk) = \sum_{n=1}^\infty \frac{1}{n!} \underset{\vk_{1\cdots n}= \vk}{\int} 
\mathcal{K}^{(n)}_{\chi\delta}(\vk_1,\cdots,\vk_n) \chi(\vk_1)\cdots\chi(\vk_n).
\end{equation}
After some iterative manipulations of Eqs.~(\ref{KGeqPert}), (\ref{KGeq_chi}), (\ref{dI}), (\ref{dIexp}) and (\ref{dchiexp}) we arrive to
\begin{equation}\label{M2}
M_2(\vk_1,\vk_2) = \frac{2 C \beta_2 A_0}{\beta^2}  \mathcal{K}^{(1)}_{\chi\delta}(\vk_1) - \frac{C^2 \kappa_3}{4 \beta^3},
\end{equation}
\begin{align}\label{M3}
M_3(\vk_1,\vk_2,\vk_3) &= \frac{3 C \beta_2 A_0}{\beta^2} \mathcal{K}^{(2)}_{\chi\delta}(\vk_1,\vk_2) \nonumber\\
&\quad - \frac{3 C^2 \beta_3 A_0}{2 \beta^3} \mathcal{K}^{(1)}_{\chi\delta}(\vk_1) + \frac{C^3 \kappa_4}{8 \beta^4},
\end{align}
with the kernels given by
\begin{align}
\mathcal{K}^{(1)}_{\chi\delta}(\vk_1) &= \frac{3}{2 A_0} \Pi(k_1), \\   
\mathcal{K}^{(2)}_{\chi\delta}(\vk_1,\vk_2) &= \frac{1}{2 A_0} M_2^\text{FL}(\vk_1,\vk_2),
\end{align}
where the frame-lagged $M_2$ function is defined in Eq.~(\ref{M2FL}).
It is convenient to symmetrize these $M$ functions over their arguments, as we do in the following.

In theories of gravity defined in the Jordan frame, $M_2$ and $M_3$ are $k$-dependent if non-canonical kinetic terms or higher derivatives of the scalar field are present in the Lagrangian; a known case with such scale dependencies is the Dvali Gabadadze Porrati (DGP) braneworld model, whereas no 
scale dependencies in the $M$s is found in f(R) Hu-Sawicki gravity; see \cite{2009PhRvD..79l3512K}.
For theories in the Einstein frame, the $k$-dependence arises due to the couplings $\chi \delta$ in the Klein-Gordon equation, even if no derivatives other than the standard kinetic term appear in their defining action.

It is worth mentioning that functions $M_2$ and $M_3$ encode the physics of particular theories, and they determine the screening properties too; 
these are the coefficients of Taylor expanding the non-linearities of the Klein-Gordon in Fourier space. As we note these functions can be positive or negative which will be responsible for the screening properties of a model.
Moreover, if $\beta_2$ and $\beta_3$ are zero, as it happens for theories with a conformal factor that is linear in the scalar field, both $M_2$ and $M_3$ become scale-independent. 
This is, for example, the case of the first proposed chameleon model (\cite{Khoury:2003aq}).

Now, we define the linear differential  operator (\cite{Matsubara:2015ipa})
\begin{equation}  
 \T = \frac{\partial^2 \,\,}{\partial t^2} + 2 H \frac{\partial \,\,}{\partial t},
\end{equation}
and the equation of motion for the displacement field divergence [Eq.~(\ref{pre_em})] becomes \citep{Aviles:2017aor}\footnote{Starting from Eq.~(\ref{pre_em}) we use 
$\nabla_{\vx \,i}(\T \Psi_i)= (J^{-1})_{ji}\nabla_{j}(\T \Psi_i)$. Afterwards we can expand $(J^{-1})_{ij} = \delta_{ij} - \Psi_{i,j} +  \Psi_{i,k}\Psi_{k,j} + \cdots$.}
\begin{align} \label{eqm}
  [(J^{-1})_{ji}\T\Psi_{i,j}](\vk) &= - A(k) \tilde{\delta}(\vk)
  + \frac{k^2/a^2}{6 \Pi(\vk)} \delta I (\vk) \nonumber \\ & + \frac{M_1}{3 \Pi(k)} \frac{1}{2a^2} [(\nabla^2_\vx \chi - \nabla^2 \chi)](\vk).
\end{align}
We perturb the displacement field as $\Ps = \lambda \Ps^{(1)} + \lambda^2 \Ps^{(2)} + \lambda^3 \Ps^{(3)} + \mathcal{O}(\lambda^4)$, and solve the above equation order by order. Stopping at third order allow us to 
calculate the first corrections to the linear power spectrum. Hereafter, we absorb the control parameter $\lambda$ in the definition of $\Ps$. To first order, Eq.~(\ref{eqm}) yields
\begin{equation} \label{em_linear}
 \big(\T - A(k)\big) (i k_i \Psi_i^{(1)}(\vk)) =0.
\end{equation}
This equation has the same form as the linear equation for the matter overdensity $\delta(\vk,t)$. Therefore, we get
\begin{equation}\label{Psi1}
 \Psi^{(1)}_i(\vk,t) = i \frac{k_i}{k^2} D_{+}(\vk,t)\tilde{\delta}_L(\vk,t_0)
\end{equation}
with $D_+(k)$ the fastest growing solution of equation $\big(\T - A(k)\big) D(k) =0$ normalized to unity as $D_+(k=0,t_0)=1$. 
This normalization is useful for theories that reduce to $\Lambda$CDM 
at very large scales, which is the case when the fifth force range is finite. 
The initial condition $\tilde{\delta}_L(\vk,t_0)$ is fixed by noting that linearizing the RHS  
of Eq.~(\ref{Jtodelta}) we have $\delta^{(1)}(\vx)= -\Psi_{i,i}^{(1)}(\vq)$. Because we are dealing with linear fields we can safely drop the tilde over the
overdensity in Eq.~(\ref{Psi1}). To second order, Eq.~(\ref{eqm}) leads to the solution 
\begin{equation}
 \label{LD2order}
\Psi^{i (2)}(\vk) = \frac{ik^i}{2k^2} \ikk \frac{3}{7} \Big( \bar{D}^{(2)}_\text{NS}(\vk_1,\vk_2) 
- \bar{D}^{(2)}_\text{S}(\vk_1,\vk_2)\Big) \delta_1\delta_2,
\end{equation}
where we denote $\delta_{1,2} \equiv  \delta_L (\vk_{1,2})$.
Momentum conservation implies $\vk =  \vk_1 + \vk_2$, as it is explicit in the Dirac delta function, cf. Eq.~(\ref{int_not}). We are splitting the second order growth in
non-screened (NS) and screening (S) pieces. These growth functions $D^{(2)}$ are solutions, with the 
appropriate initial conditions, to the equations
\begin{align}
  \big(\T - A(k)\big) D^{(2)}_\text{NS}(\vk_1,\vk_2) &=    \Bigg( A(k) - (A(k_1) + A(k_2) - A(k) ) \frac{(\vk_1 \cdot \vk_2)^2}{k_1^2 k_2^2}  \nonumber\\
 +(A(k)-A(k_1))\frac{\vk_1\cdot\vk_2}{k_2^2} & + (A(k)-A(k_2))\frac{\vk_1\cdot\vk_2}{k_1^2} \Bigg)D_+(k_1) D_+(k_2) \label{D2NoSc} \\
 \big(\T - A(k)\big) D^{(2)}_\text{S}(\vk_1,\vk_2) &= 
 \left(\frac{2 A_0}{3}\right)^2 \frac{k^2}{a^2}\frac{M_2(\vk_1,\vk_2) D_+(k_1) D_+(k_2)}{6 \Pi(k)\Pi(k_1)\Pi(k_2)} \nonumber\\
  &\equiv \mathcal{S}^{(2)}_\text{S}, \label{D2dI} 
\end{align}
and the normalized growth functions are defined as
\begin{align}
 \bar{D}^{(2)}_{\text{S},\text{NS}}(\vk_1,\vk_2,t) &= \frac{7}{3}\frac{D^{(2)}_{\text{S},\text{NS}}(\vk_1,\vk_2,t) }{D_+(k_1)D_+(k_2)}. 
\end{align}
In an EdS universe one obtains the well known result $\bar{D}^{(2)}_{\text{NS}}= 1 -(\hat{\vk}_1\cdot\hat{\vk}_2)^2$, while in $\Lambda$CDM one gets the same result multiplied by a function that varies slowly with time, such that nowadays  $\bar{D}^{(2)\Lambda\text{CDM}}_{\text{NS}} \simeq 1.01 \bar{D}^{(2)\text{EdS}}_{\text{NS}}$. 
The screening second order growth, $\bar{D}^{(2)}_{\text{S}}$, is zero in both EdS and $\Lambda$CDM models. 

The function $\bar{D}^{(2)}_\text{S}$ will be important for our discussion. It encodes the non-linearities of the ``potential'' of the scalar field and it yields the second order screening effects that drive
the theory to GR at small scales. The total second order growth function, as can be read from Eq. (\ref{LD2order}), 
is given by $D^{(2)} = D^{(2)}_\text{NS} - D^{(2)}_\text{S}$, such that negative values of $D^{(2)}_\text{S}$ enhance the growth of perturbations ({\it anti-screening effects}) 
while positive values of it yield the standard suppression of the fifth force.

Analogously, each higher perturbative order carries its own screening and it is efficient over a certain $k$ interval. 
The third order Lagrangian displacement field can be computed to 
give
\begin{equation} \label{LD3order}
\Psi^{(3)}_i(\vk) = \frac{ik_i}{6k^2}\ikkk
\bar{D}^{(3)}(\vk_1,\vk_2\vk_3)
\delta_1\delta_2\delta_3.
\end{equation}
with normalized growth
\begin{equation}
\bar{D}^{(3)}(\vk_1,\vk_2\vk_3) = \frac{D^{(3)}(\vk_1,\vk_2\vk_3)}{D_+(k_1)D_+(k_2)D_+(k_3)}
\end{equation}
The complete expression for the third order growth function $D^{(3)}$, equivalent to Eq.~(\ref{LD2order}), is large and can be found in \cite{Aviles:2017aor}, though in  
Eq.~(\ref{D3symm}) below we show this function for a particular configuration of wavevectors. In sec.\ref{sec:Growth} we will be interested in splitting the growth in non-screened and screening pieces.
Unlike the second order case, at third order the decomposition cannot be performed directly through the linear differential equations that govern the growth. Hence, in this case  
$\bar{D}^{(3)}_\text{NS}$ is obtained by setting $M_2$ and $M_3$ equal to zero,
\begin{equation}
\bar{D}^{(3)}_\text{NS} \equiv  \bar{D}^{(3)}|_{M_2=M_3=0},    
\end{equation}
while $\bar{D}^{(3)}_\text{S}$, by the relation
\begin{equation} \label{D3Sdef}
 \bar{D}^{(3)} =  \bar{D}^{(3)}_\text{NS}  -\bar{D}^{(3)}_\text{S}.
\end{equation}
In such a way third order screenings are realized by having $\bar{D}^{(3)}_\text{S}>0$,  while anti-screening by $\bar{D}^{(3)}_\text{S}<0$.

\end{section}

\begin{section}{Modified gravity theories with different screening mechanisms}\label{sec:MGmodels}
As shown above, expressions $M_2(\vk_1,\vk_2)$ and $M_3(\vk_1,\vk_2)$ depend upon the explicit form of the 
Klein-Gordon equation, that in turn depends on the frame posed. The formalism developed 
here applies to the Einstein frame in which the coupling function $C(\varphi)$ is nontrivial, but  applies also to the Jordan frame by setting $C(\bar{\varphi})\equiv 1$, as explicitly done in \cite{Aviles:2017aor}. 

In the following we treat examples 
of models with different screening properties. We start with Symmetrons that are posed in the Einstein frame and follow with f(R) Hu-Sawicki and DGP models posed in the Jordan frame.

\begin{subsection}{Symmetron model}
 
The Symmetron model can be introduced with the action of Eq.~(\ref{EF_action}) with a self-interacting potential
\begin{align}
 V(\varphi) &= V_0 -\frac{1}{2}\mu^2 \varphi^2 + \frac{1}{4}\lambda \varphi^4,  
 \end{align}
and the conformal factor
\begin{align}
  C(\varphi) &= 1 + \frac{1}{4}\left(\frac{\varphi}{M} \right)^2.
\end{align}
Assuming the background piece of the scalar field is always sitting in the minimum of the effective potential $V_{eff}=V + \bar{\rho} C$,
one obtains 
\begin{equation}
 \bvarphi = \bvarphi_0  \sqrt{ 1 - \left(\frac{a_{ssb}}{a}\right)^3},
\end{equation}
where $a_{ssb}$ is the scale factor at which the $\mathbb{Z}_2$ symmetry is broken. 
The scalar field effective mass  and the strength $\beta$ of the fifth force are
\begin{align}
 m(a) &=  m_0 \sqrt{ 1 - \left(\frac{a_{ssb}}{a}\right)^3}, \qquad
 \beta(a) = \beta_0\sqrt{ 1 - \left(\frac{a_{ssb}}{a}\right)^3}. 
\end{align}
These functions are commonly generalized to
\begin{align}
 m(a) &=  m_0 \left[ 1 - \left(\frac{a_{ssb}}{a}\right)^3 \right]^{\hat{m}},\qquad
 \beta(a) = \beta_0 \left[ 1 - \left(\frac{a_{ssb}}{a}\right)^3 \right]^{\hat{n}}. 
\end{align}
In this way a Symmetron model can be characterized by the set of parameters $(a_{ssb},m_0,\beta_0,\hat{n},\hat{m})$. Although, other equivalent parameters are also used in the literature; for example, in order to contain the parameters $\mu$, $\lambda$ and $M$, instead.
Since variations of fermion masses cannot vary too much over the Universe lifetime, we simply set $C(\bvarphi)=1$.

The function $M_1$ plays an important role in the upcoming discussion. It is given by
\begin{equation}
 M_1(a) = \frac{C  m(a)^2}{2\beta(a)^2} = \frac{C m_0^2}{2 \beta^2_0} \left[ 1 - \left(\frac{a_{ssb}}{a}\right)^3 \right]^{2(\hat{m}-\hat{n})}.
\end{equation}
We soon notice that if $\hat{m}=\hat{n}$, $M_1$ becomes a constant. 
Expressions for $M_2$ and $M_3$, given by Eqs.~(\ref{M2}) and (\ref{M3}), depend on the conformal coupling, $\beta_{n}$, $\kappa_n$, and $\mathcal{K}^{(n)}_{\chi\delta}$. In the present case these formulae are cumbersome and we do not show here, but we solved them numerically when integrating the differential equations for the growth functions and to construct power spectra.   $M_2$ and $M_3$ are indeed important to determine the fate of nonlinearities. Cosmological screenings are encoded in these functions and hence they serve to distinguish among different screening types. In the present case, $M_2$ and $M_3$ turn out to be negative for certain cosmological epoch and specific wavenumbers that will be reflected in an anti-screening effect in the power spectra shown in next section.

\end{subsection}

\begin{subsection}{f(R) Hu-Sawicki model - Chameleon mechanism}

Here we consider a Lagrangian density given by  ${\cal L} = \frac{1}{2} \Mp^{2}  \sqrt{-g}(R + f(R))$, in contrast to Eq.~(\ref{EF_action}). As it is known, f(R)
models can be brought to a Jordan frame description and one then can apply our perturbation formalism \citep{Aviles:2017aor}. We analyze the Hu-Sawicki model with parameter $n=1$. 
One can define the scalar degree of freedom to be $\chi = \delta f_R$, with $f_R=df/dR$, for which one finds a Klein-Gordon equation, that in the quasi-static limit is 
\begin{equation} \label{KG_chi_fR}
 \frac{3}{a^2}\nabla^2_\vx \chi = -2 A_0  \delta + \delta R,
\end{equation}
where $\delta R = R(f_R) - R(\bar{f}_R) = M_1 \chi + \frac{1}{2} M_2 \chi^2 +\frac{1}{6} M_3 \chi^3 +\cdots $.  By developing Eq.~(\ref{KG_chi_fR}) in terms of the scalar field $\chi$, 
and setting $C=1$, one arrives to Eq.~(\ref{KGeq_chi}) with $2\beta^2=1/3$. The $M$ functions are obtained by using $R(f_{R})\simeq \bar{R} (f_{R0}/f_R)^{1/2}$  and  
$\bar{R}=3H_0^2(\Omega_{m0}a^{-3} + 4 \Omega_\Lambda)$: 
\begin{align}\label{M1fR}
M_1(a) = \frac{3}{2}  \frac{H_0^2}{|f_{R0}|} \frac{(\Omega_{m0} a^{-3} + 4 \Omega_\Lambda)^3}{(\Omega_{m0}  + 4 \Omega_\Lambda)^2}, 
\end{align}
\begin{align}\label{M2fR}
M_2(a) = \frac{9}{4}  \frac{H_0^2}{|f_{R0}|^2} \frac{(\Omega_{m0} a^{-3} + 4 \Omega_\Lambda)^5}{(\Omega_{m0}  + 4 \Omega_\Lambda)^4}, 
\end{align}
\begin{align}\label{M3fR}
M_3(a)=  \frac{45}{8}  \frac{H_0^2}{|f_{R0}|^3} \frac{(\Omega_{m0} a^{-3} + 4 \Omega_\Lambda)^7}{(\Omega_{m0}  + 4 \Omega_\Lambda)^6}, 
\end{align}
while the scalar field mass is given by $m=\sqrt{M_1/3}$.
We will use values $f_{R0}=-10^{-4}, -10^{-8}, -10^{-12}$ that correspond to models F4, F8, and 
F12, respectively. The fact that for these values of $f_{R0}$ the expansion history is 
indistinguishable from that in $\Lambda$CDM, as we have assumed, has been studied in \cite{Hu:2007nk}.

Note that $M_2$ and $M_3$, that determine the screening properties, result to be $k-$independent and positive. This is a non-trivial feature since Eqs.~(\ref{M2}) and (\ref{M3}) may also be negative, as we saw for the Symmetron. In the f(R) case, being the $M$s positive, implies a normal screening (in opposition to what we found in the Symmetron case: \textit{anti-screening}). 

\end{subsection}

\begin{subsection}{Cubic Galileons and DGP models - Vainshtein mechanism }

Cubic Galileons (\cite{2009PhRvD..79f4036N}) and DGP (\cite{2000PhLB..485..208D}) models stem from different physical motivations, with different background dynamics,  but they share a similar  structure. Both are theories defined in the Jordan frame with Klein-Gordon equation in the quasi-static limit given by 
\begin{equation} \label{KGDGP}
 \frac{1}{a^2}\nabla^2_\vx \chi = - Z_1(a) \big( (\nabla_\vx^2 \chi)^2 - (\nabla_{\vx\,i}\nabla_{\vx\,j} \chi)^2 \big) - Z_2(a) \rho
\end{equation}
where $Z_1$ and $Z_2$ are model dependent functions of time. We note the scalar field becomes massless, such that $M_1=0$ and therefore the linear growth $D_+$ is scale independent.
The screening on these models is provided by 
the Vainshtein mechanism, that arise from the 
nonlinear derivatives terms in the Klein-Gordon equation.
The functions $M_2$ and $M_3$ become (\cite{Aviles:2018saf})
\begin{align}
 M_2(\vk_1,\vk_2) &=  \frac{Z_1(a)}{\beta^2}  \left[ k_1^2 k_2^2 - (\vk_1 \cdot \vk_2)^2 \right]  \label{M2DGP} \\
 M_3(\vk_1,\vk_2,\vk_3) &= \frac{3 Z_1(a)}{2 a^2 \beta^4 A_0}  \Big[ 
   2 (\vk_1 \cdot \vk_2)^2 k_3^2 + (\vk_1 \cdot \vk_2 ) k_1^2 k_3^2  \nonumber\\
   - (\vk_1 \cdot &\vk_2)(\vk_1\cdot\vk_3)^2 - 2(\vk_1\cdot\vk_2)(\vk_2\cdot\vk_3)(\vk_3\cdot\vk_1) \Big].  \label{M3DGP}
\end{align}
The function $M_3$ appears when transforming from Eulerian to Lagrangian derivatives in Eq.~(\ref{KGDGP}).  Given the above structure the screenings in both cubic Galileon and DGP are similar. Functions $\beta$, $Z_1$, and $Z_2$ depend on the specific model. For definiteness we consider here DGP: 
\begin{align}
\beta^2&= \frac{1}{6}\left[1 - 2 \epsilon H r_c \left(1+\frac{\dot{H}}{3 H^2} \right)\right]^{-1}, \\
Z_1 &= \frac{2 \beta^2 r^2_c}{a^4}, \\
Z_2 &= 4 A_0 \beta^2,
\end{align}
where $\epsilon=+1$ for the self-acceleration branch and $-1$ for the normal branch. $r_c$ is the crossover scale below that the theory behaves as a scalar tensor theory.
For a side by side comparison of both models and for the expression of the functions $\beta$, $Z_1$, and $Z_2$ in cubic Galileons see \cite{2013JCAP...10..027B}.

\end{subsection}

\end{section}

\begin{section}{Matter power spectrum}\label{sec:PS}

\begin{figure*}
   \centering
   \includegraphics[width=3.2 in]{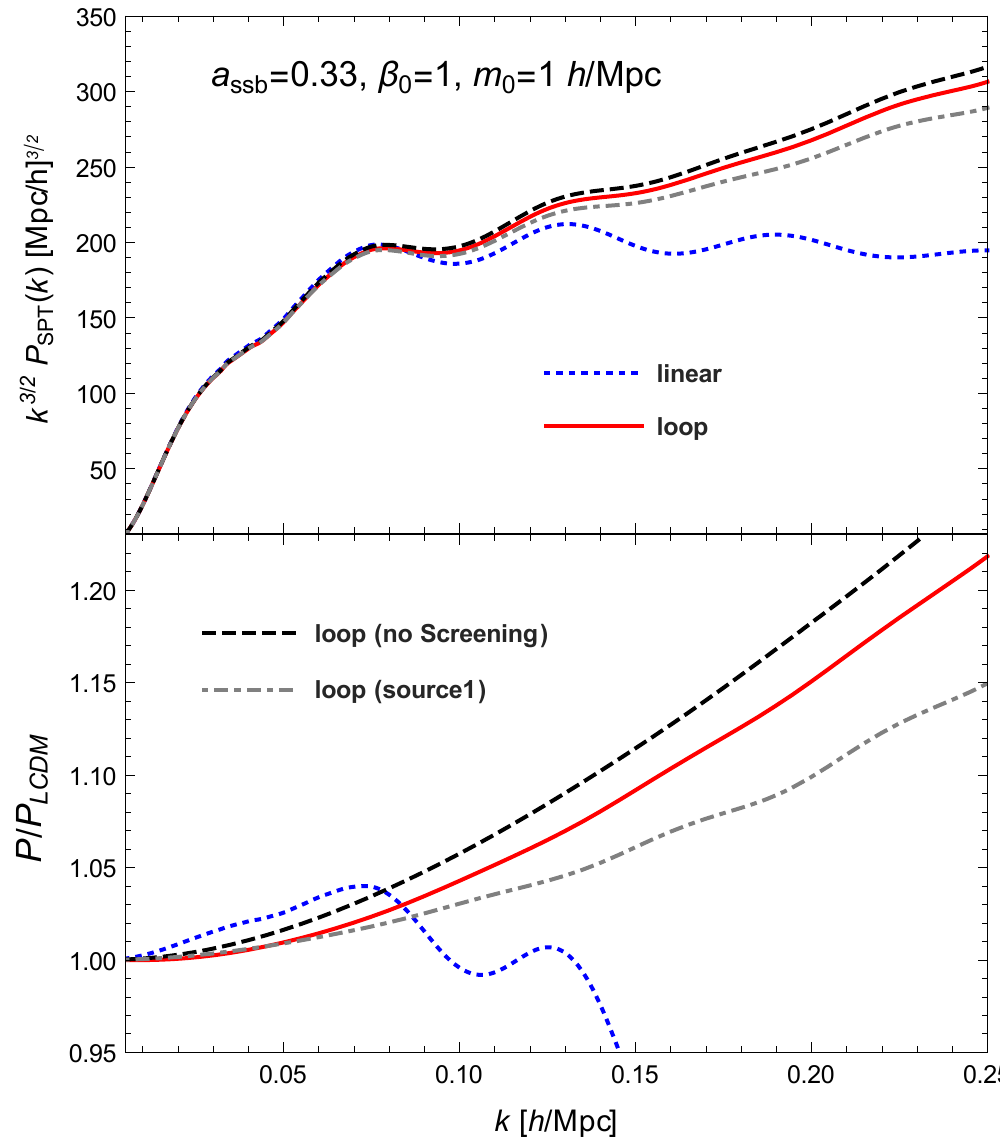}
   \includegraphics[width=3.2 in]{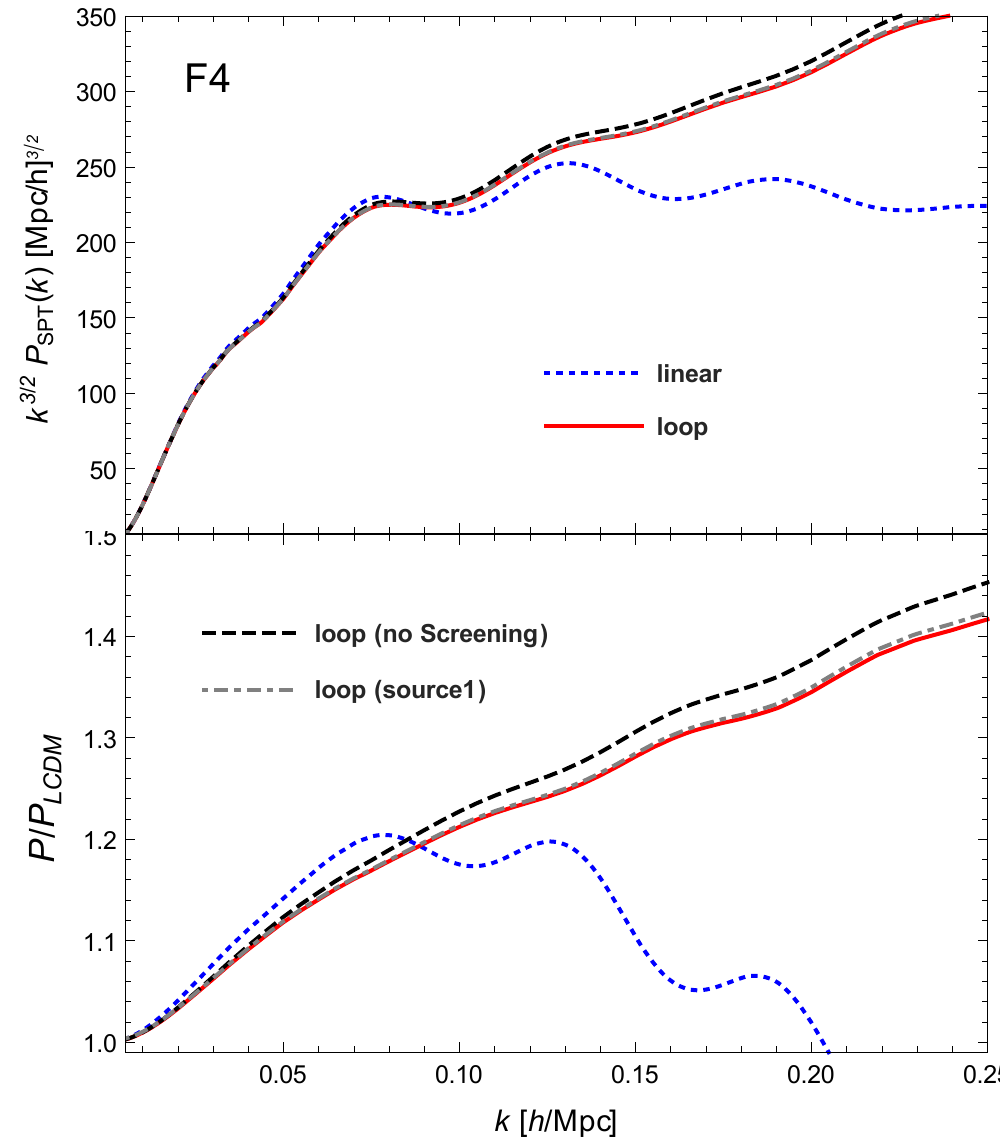}   
   \caption{SPT power spectra for Symmetron model (left panel) with $a_{ssb}=0.33$, 
   $m_0=1 h/\text{Mpc}$ and $\beta_0=1$, and F4 model (right panel), 
   evaluated at redshift $z=0$. We plot the linear theory, the full SPT,
    the SPT without screenings, and the SPT considering only $S_1$ (source1) screenings (see text for details). 
	In the lower panels we take their ratios to the 
	1-loop SPT power spectrum for $\Lambda$CDM model.}
              \label{fig:symmetronPS}%
    \end{figure*}

For MG models with an early EdS phase, as we posit here, the linear matter power spectrum is given by
\begin{equation}
 P_L(k,t) = (D_+(k,t))^2 P_L^{\Lambda\text{CDM}}(k,t_0). 
\end{equation}
The building blocks for loop matter statistics are the functions
\begin{align}
 Q_1(k) &=  \int \Dk{p} 
\left( \bar{D}^{(2)}_\text{NS}  
- \bar{D}^{(2)}_\text{S} \right)^2  P_L(p)P_L(|\vk-\vp|), \label{Q1f} \\ 
 Q_2(k) &=\int \Dk{p} \frac{\vk\cdot\vp}{p^2}\frac{\vk\cdot(\vk-\vp)}{|\vk-\vp|^2} 
\Bigg( \bar{D}^{(2)}_\text{NS}  -  \bar{D}^{(2)}_\text{S} \Bigg) P_L(p)P_L(|\vk-\vp|),  \label{Q2f} \\ 
 Q_3(k) &= \int \Dk{p} \frac{(\vk\cdot\vp)^2}{p^4}\frac{(\vk\cdot(\vk-\vp))^2}{|\vk-\vp|^4} P_L(p)P_L(|\vk-\vp|), \label{Q3f} \\ 
 R_1(k) &= \int \Dk{p} \frac{21}{10} \frac{D^{(3)s}(\vk,-\vp,\vp)}{D_{+}(k)D_{+}^2(p)} P_L(p)P_L(k), \label{R1f} \\ 
  R_2(k) &= \int \Dk{p} \frac{\vk\cdot\vp}{p^2}\frac{\vk\cdot(\vk-\vp)}{|\vk-\vp|^2} 
\Bigg( \bar{D}^{(2)}_\text{NS}-  \bar{D}^{(2)}_\text{S} \Bigg) P_L(p)P_L(k), \label{R2f}
\end{align}
where the normalized growth functions in Eqs.~(\ref{Q1f}) and (\ref{Q2f}) are evaluated as $D^{(2)}(\vp,\vk-\vp)$, while in Eq.~(\ref{R2f}) as
$D^{(2)}(\vk,-\vp)$. These functions were introduced in (\cite{Mat08a}) for EdS evolution, and extended in (\cite{Aviles:2017aor}) for MG. From them,
power spectra and correlation functions in different resummation schemes can be obtained; see, for example (\cite{Carlson:2012bu,Matsubara:2008wx,Vlah:2015sea}). In particular, the SPT power spectrum is defined as
\begin{equation}
 P_\text{SPT}(k) = P_L + P_{22} + P_{13}
\end{equation}
and it can be shown that for $\Lambda$CDM (\cite{Mat08a,Vlah:2014nta}) and for MG (\cite{Aviles:2018saf}) the following expressions hold:
\begin{align}
P_{22}(k) &= \frac{9}{98}Q_1(k) + \frac{3}{7}Q_2(k) + \frac{1}{2}Q_3(k)  \\
P_{13}(k) &= \frac{10}{21}R_1(k) + \frac{6}{7}R_2(k) - \sigma^2_L k^2 P_L(k), 
\end{align}
where the one dimensional variance of linear displacement fields is
\begin{equation}
 \sigma^2_L = \frac{1}{3}\delta_{ij} \langle \Psi_i(0)\Psi_j(0) \rangle = \frac{1}{6 \pi^2} \int_0^\infty dp P_L(p).
\end{equation}
In Eq.~(\ref{R1f}) the third order growth function $D^{(3)s}$ is the solution to
\begin{equation} \label{D3symm}
 D^{(3)s}(\vk,-\vp,\vp) = (\T -A(k))^{-1} \Big( \mathcal{S}_1 + \mathcal{S}_2 + \mathcal{S}_3 \Big)
\end{equation}
where the label ``$s$'' means that $D^{(3)}(\vk_1,\vk_2,\vk_3)$ is symmetrized over its arguments; afterwards it is evaluated in the double squeezed configuration 
given by $\vk_3=-\vk_2 = \vp$ and $\vk_1 = \vk$. We note from Eq.~(\ref{R1f}) that these are the only quadrilateral configurations ---subject to momentum conservation: 
$\vk = \vk_1 + \vk_2 + \vk_3$--- that survive in the 1-loop computations. 
Expressions for the sources $\mathcal{S}_i$ are written in appendix \ref{app:3rdOrder}.
We have split the sources since in $\Lambda$CDM only $\mathcal{S}_1$ is present, and in general it is the dominant contribution. Source $\mathcal{S}_2$ is a mix of frame-lagging and terms coming from the scale dependent gravitational strength, that yields a small contribution. Meanwhile,
$\mathcal{S}_3$ is only composed by screenings; indeed, this is the only source containing the function $M_3$.

In left panel of Fig.~\ref{fig:symmetronPS} we plot power spectra for a Symmetron model with $a_{ssb}=0.33$,
$m_0=1 \, h/\text{Mpc}$ and $\beta_0=1$, and $\hat{n}=\hat{m}=0.5$. We do it for the following cases: full loop SPT (solid red curves); without screenings (dashed black curves); 
considering only source $\mathcal{S}_1$ in Eq.~(\ref{D3symm}) (dot-dashed gray curves); and
the linear power spectrum (dotted blue curves). The lower panel in this figure shows the ratios of the different power spectra to the 1-loop $\Lambda$CDM power spectra, for which we assumed the reference cosmology as given by WMAP 2009 best fits ($\Omega_m = 0.281$, $\Omega_b=0.046$, $h = 0.697$, and $\sigma_8=0.82$). 
An interesting observation from these plots is that the perturbative screenings do not always act in the ``screening direction''; that is, although the power spectrum with only source $\mathcal{S}_1$ 
has less screening contributions than the full loop curves, it is actually closer to the GR power spectrum.\footnote{ Keeping only $\mathcal{S}_1$ is equivalent (up to frame-lagging terms) to keep only the $\gamma_2$ term in the expression of $\delta I$ of 
\cite{Bose:2016qun} that Fourier expands it in powers of matter overdensities, instead of in powers of the scalar field as we do in Eq. (\ref{dIexp}).} 
We interpret this fact as 
the non-linearities of Klein-Gordon equation due to the couplings $\chi \delta$ also provide anti-screening effects. 
This behaviour can be observed in the figures for Symmetron models in \cite{Brax:2013fna}, but unfortunately it is not discussed in that paper. 
An analogous plot for the F4 chameleon model is shown in the right panel of Fig.~\ref{fig:symmetronPS}, from here we notice that the different 
screening contributions always act in the same ``direction'', driving the theory to GR; we further notice that in this case the contributions due to the sources $\mathcal{S}_2$ and $\mathcal{S}_3$ are almost negligible. 
In the following section we will discuss these effects
in more detail by considering the functions $\bar{D}^{(2)}_\text{S}$ and $\bar{D}^{(3)}_\text{S}$ for the Symmetron, defined in the Einstein frame, and for the DGP and f(R) models that are defined in the Jordan frame.

\end{section}

\begin{section}{Screening growth functions}\label{sec:Growth}
 
In this section we study the main features of the normalized second and third order growth functions $\bar{D}^{(2)}_\text{S}(\vk_1,\vk_2)$ and 
$\bar{D}^{(3)}_\text{S}(\vk_1,\vk_2,\vk_3)$.

Since there is a Dirac delta function accompanying the second order growth functions and ensuring that $\vk = \vk_1 + \vk_2$, these three 
wavevectors form a triangle. Therefore, by assuming statistical homogeneity and isotropy, the growth functions depend only on three positive numbers, e.g. the lengths of the sides of the triangles, thus we can write
$\bar{D}^{(2)}_\text{S}(\vk_1,\vk_2) = \bar{D}^{(2)}_\text{S}(k,k_1,k_2)$. Three triangle configurations will be considered: equilateral, $k = k_1= k_2$; orthogonal
$k_1 = k_2 = \sqrt{2} k$; and squeezed, $k \simeq k_1$, $k_2\simeq 0$. There is a second squeezed configuration with $k\simeq 0$ and  $k_1 \simeq k_2$ 
corresponding  to very large scales, where the fifth force vanishes for massless theories and the screenings are zero. Analogously, we will consider third order screening growths given by Eq.~(\ref{D3Sdef}), in the doubled squeezed configuration explained above.


We stress out that $\bar{D}^{(2)}_\text{S}$ and $\bar{D}^{(3)}_\text{S}$ with positive values will screen the fifth force while negative values will anti-screening it instead. Given this, the $M$ functions of each model  will determine their screening properties. 

\subsection{Chameleon screening in the Jordan Frame - the f(R) case}

\begin{figure}
	\begin{center}
	\includegraphics[width=3.2 in]{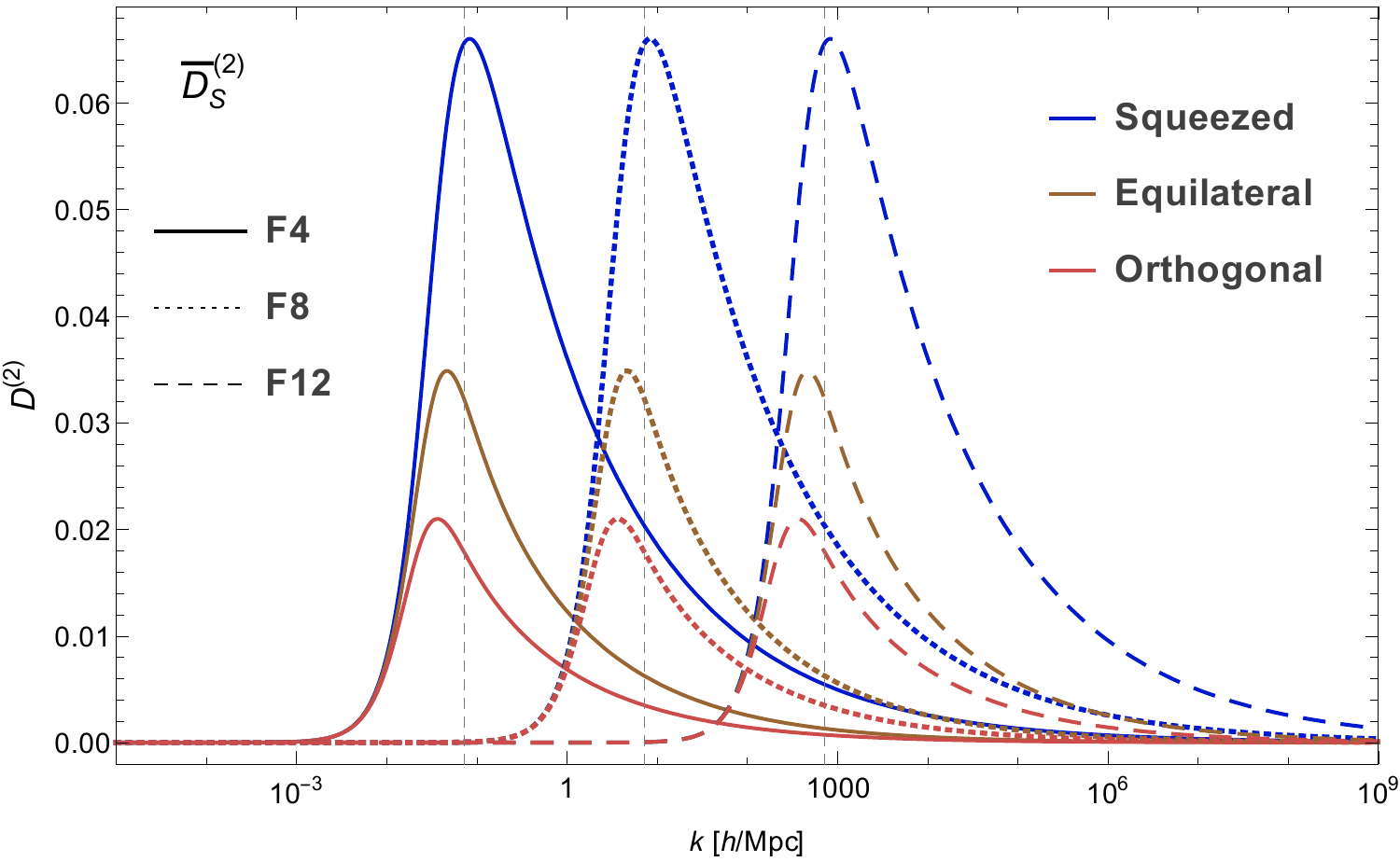}
	\includegraphics[width=3.2 in]{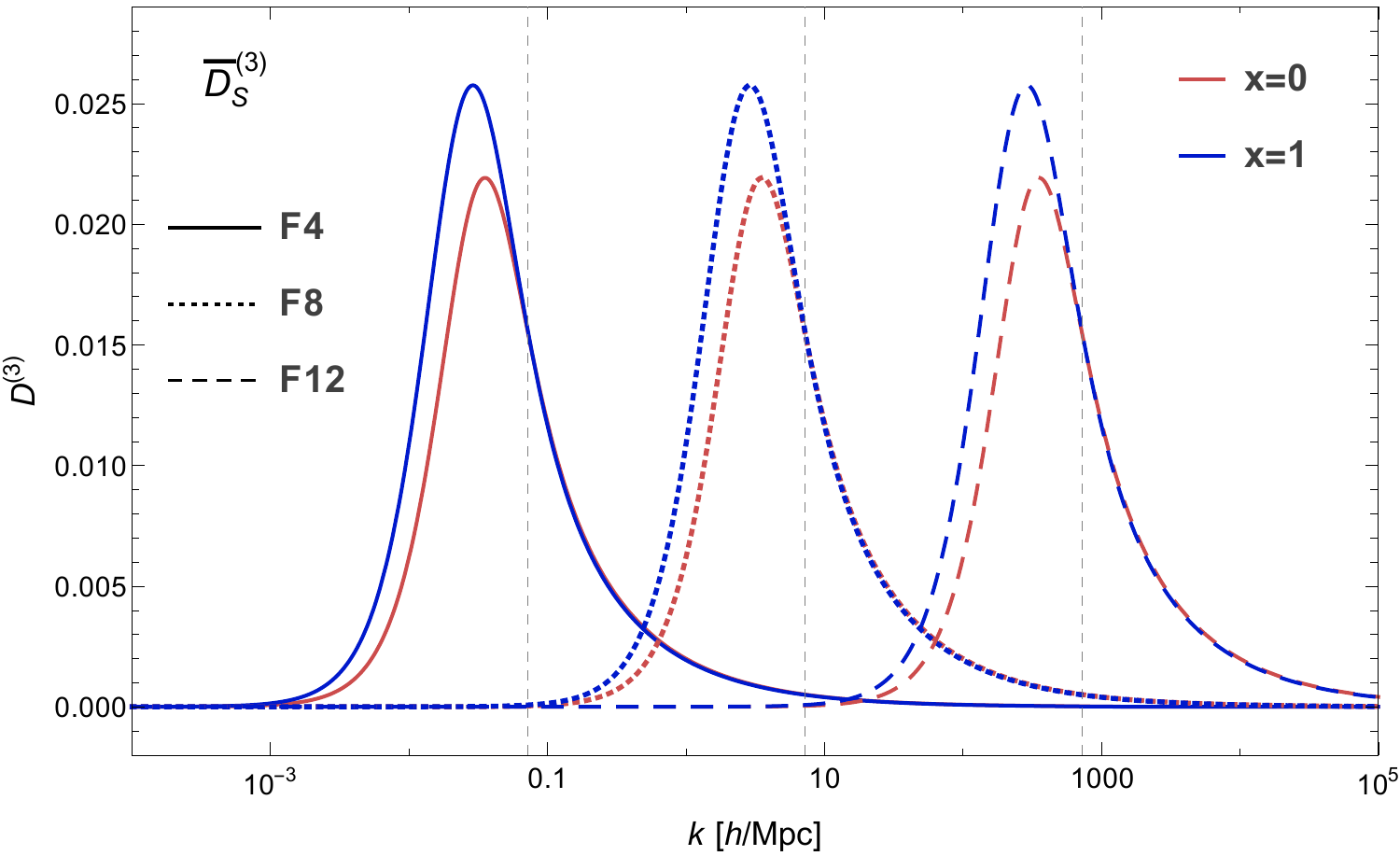}
	\caption{$ \bar{D}^{(2)}_\text{S}$: contributions to the second order normalized growth functions due to screenings for F4, F8 and F12 models (upper panel), considering different triangle configurations. The vertical lines are located at $k=a \sqrt{M_1}$, showing
	the characteristic scale at which the screenings are present. The lower panel
	shows the third order growth $ \bar{D}^{(3)}_\text{S}$ for different angles $x=\hat{\vk}\cdot\hat{\vp}$ and with
	$k=p$.
	\label{fig:F4F8F12}}
	\end{center}
\end{figure}

First, we consider the 
f(R) Hu-Sawicki model for different $f_{R0} = -10^{-4},-10^{-8},-10^{-12}$ 
corresponding to F4, F8, and F12 models. In the upper panel of Fig.~\ref{fig:F4F8F12} we show plots for the second order screening growth functions, $\bar{D}^{(2)}_\text{S}$, evaluated at redshift $z=0$ and for different 
triangular configurations. The vertical lines correspond to the {\it screening wavenumber}
\begin{equation}
k_{M_1} \equiv a \sqrt{M_1(a)},  
\end{equation}
which provides us with a rough estimation to characterize the scale at which the screening is present; 
in fact it is close to the maximum screening growth of the largest triangular contribution (squeezed modes). Note that in f(R) the screening scale has a simple dependence 
$k_{M_1} \propto \sqrt{1/|f_{R0}|}$, as can be seen from Eq.~(\ref{M1fR}).  The lower panel of Fig.~\ref{fig:F4F8F12} shows the 
growth $D^{(3)}_{S}(\vk,-\vp,\vp)$ for the double squeezed configuration with
additionally $|\vk|=|\vp|$ and for different values of the cosine angle $x=\hat{\vk} \cdot \hat{\vp}$.


 $k_{M_1}$ should be understood as a phenomenological scale that do not show where the screening effects start to be present, but its
usefulness stems from that it characterizes the screening for the models considered in this work. To show this at different redshifts, we
study the effects of the background evolution on the screening growth, in Fig.~\ref{fig:F4zs} we show it for the F4 model
at different redshifts $z=0,3,$ and $10$. The upper panel uses a $\Lambda$CDM background cosmology and the lower panel an EdS background evolution. 
In $\Lambda$CDM the screening curves are narrower and reach smaller maxima, this is expected 
because the cosmic acceleration attenuates the clustering of dark matter, and hence the nonlinear effects. 
Instead, in an EdS background the pattern of the growth is preserved and only shifted by the scale $k_{M_1}$ (denoted again by the vertical lines). 
Though this scale is useful, we notice that it does not provide the range over which the screening growth is present. 
This will be more evident in the following section where we study the Symmetron model and show that the behavior of the growth will be very different below and above $k_{M_1}$.

It is worthy to observe that in f(R) the nonlinearities of the Klein-Gordon equation lead to screening for any configuration. This is manifested in  
Figs.~\ref{fig:F4F8F12} and \ref{fig:F4zs} because the screening growth functions always take positive values. 
To remark this behaviour, in Fig.~\ref{fig:F4D2compare} we plot the second order normalized growth functions $\bar{D}^{(2)}$ and $\bar{D}^{(2)}_\text{S}$ (upper panel) and their ratios (bottom panel) for the equilateral
and orthogonal configurations. Because the ratios always take smaller values than unity, the standard screening behavior is always accomplished. This a consequence of Eq.~(\ref{M2fR}) which exhibits that $M_2$ is always positive. 
Below we show that this is not necessary the case for models defined in the Einstein frame.


\begin{figure}
	\begin{center}
	\includegraphics[width=3.2 in]{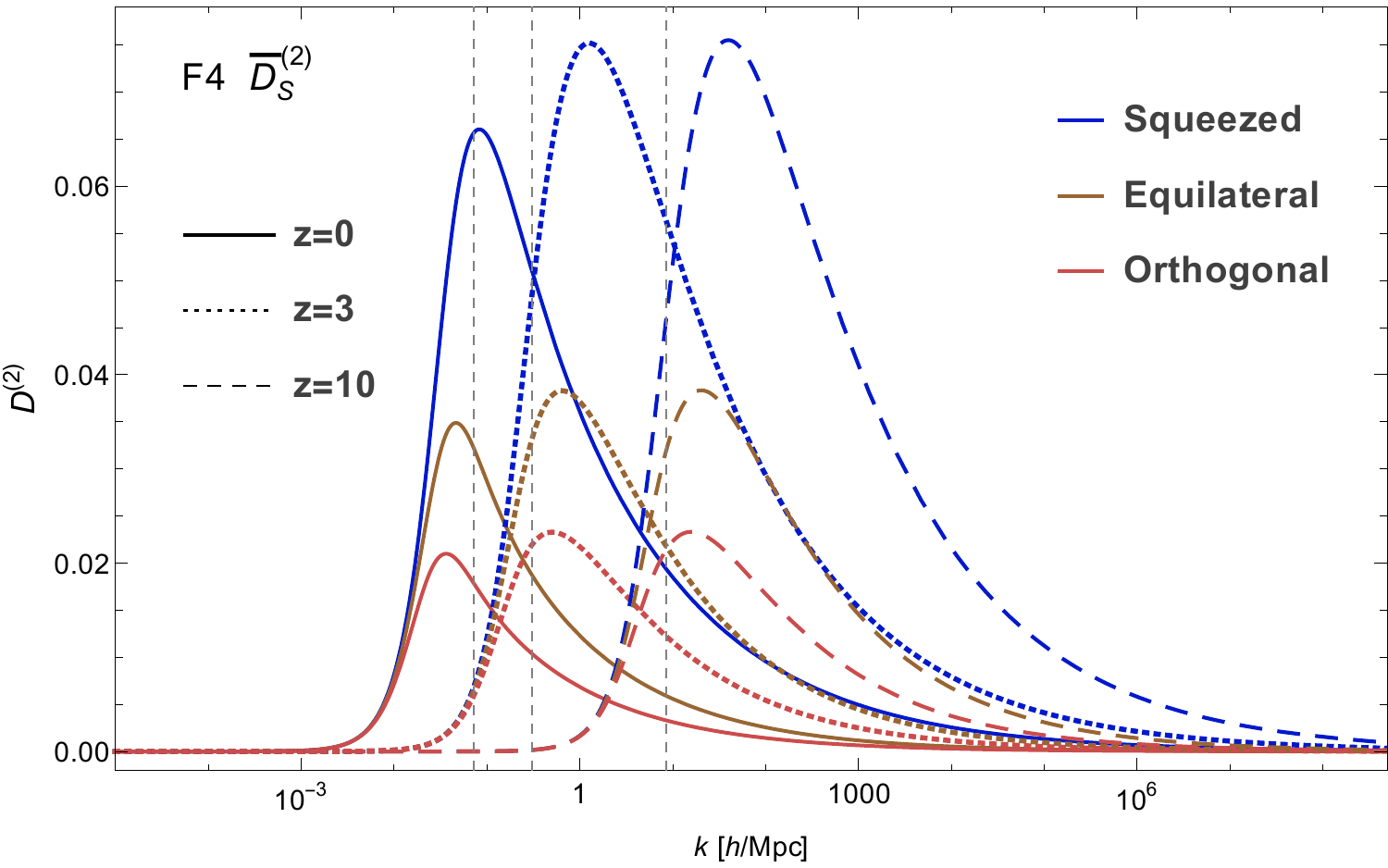}
	\includegraphics[width=3.2 in]{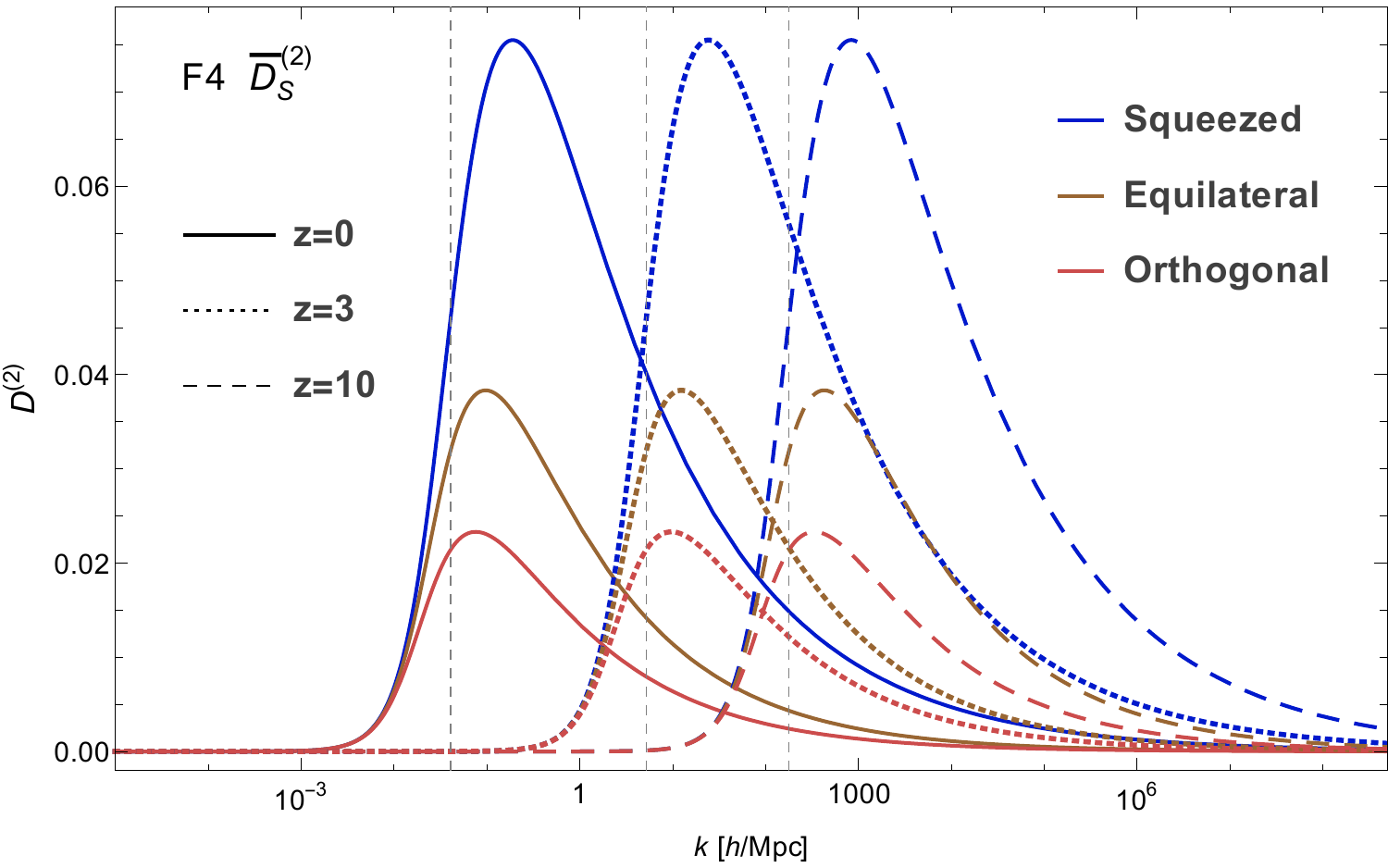}
	\caption{$ \bar{D}^{(2)}_\text{S}$ functions for the F4 for redshifts $z=0,3,10$, and considering different triangle configurations. The vertical lines are located at $k=a \sqrt{M_1}$, showing
	the characteristic scale at which the screenings are present. The upper panel shows a $\Lambda$CDM background evolution while the bottom panel an EdS background evolution.
	\label{fig:F4zs}}
	\end{center}
\end{figure}

\begin{figure}
	\begin{center}
	\includegraphics[width=3.4 in]{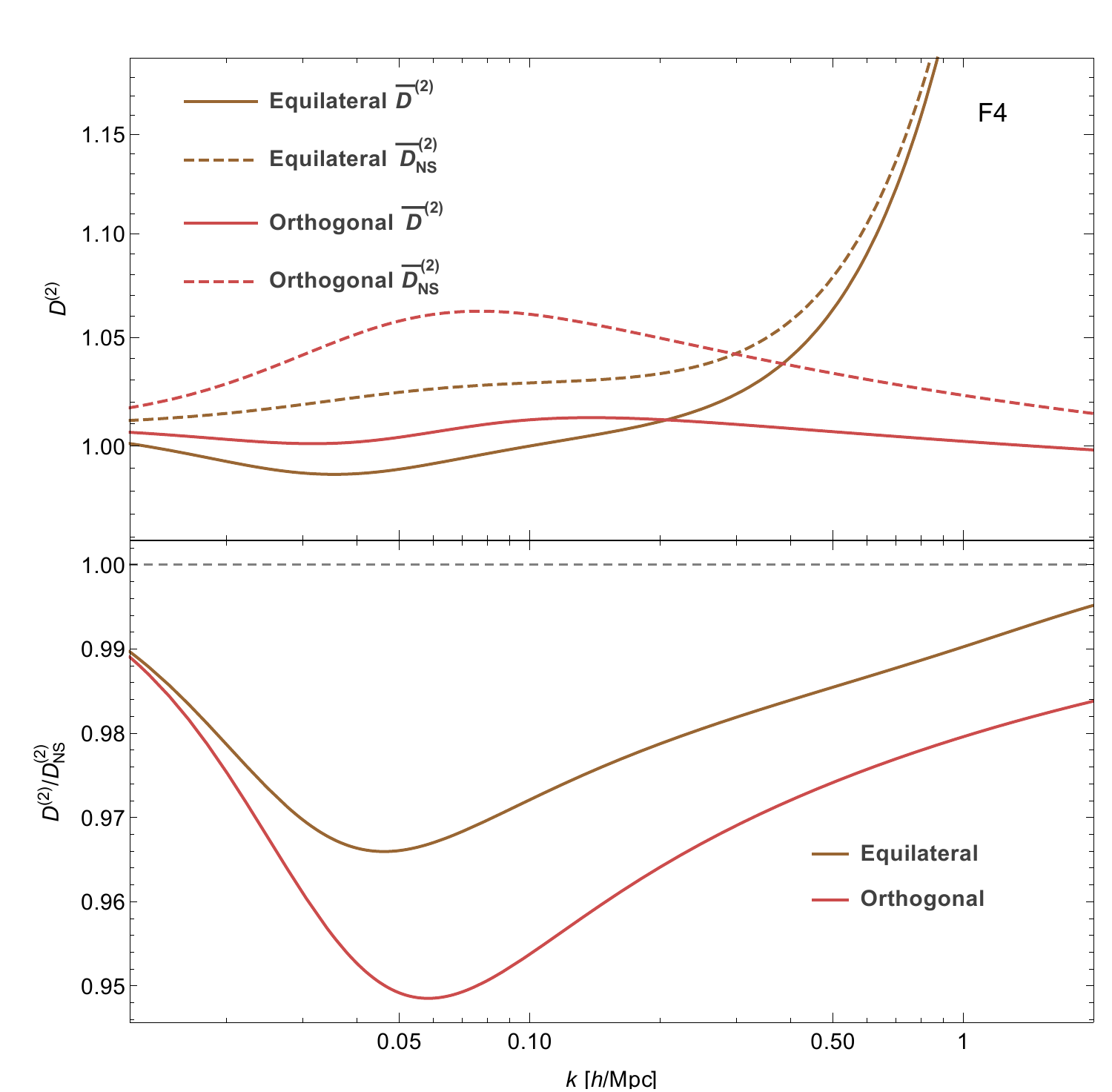}
	\caption{Normalized second order growth functions $ \bar{D}^{(2)}$ (solid curves) and $ \bar{D}^{(2)}_{NS}$ (dashed curves) for F4 model 
	at redshift $z=0$, and considering different equilateral (brown) and orthogonal (red) configurations. 
	The bottom panel shows the ratio between them. Since these take values less than 1, the growth is always suppressed in the screened $ \bar{D}^{(2)}$ case. 
	\label{fig:F4D2compare}}
	\end{center}
\end{figure}

\subsection{Symmetron screening}

Now consider the case of Symmetrons. In Fig.~\ref{fig:symmassb033} we show the second order screening growth functions for models with $\hat{n}=\hat{m}=1/2$, $a_{ssb}=0.33$, and different values of $m_0$ and
$\beta_0$. The vertical lines correspond again to the characteristic scale $k_{M_1}$. We are interested to study also the growth functions at different redshifts, 
but since for $\hat{n}=\hat{m}$ the function $M_1$ becomes a constant, the corresponding $k_{M_1}$ values are only rescaled by the scale factor. For this reason we do it with a model with
$\hat{n}=0.25$ and $\hat{m} = 0.5$, the other parameters are fixed to $a_{ssb}=0.33$, $m_0=1 h/\text{Mpc}$, and $\beta_0=1$,  for this case we show plots in Fig.~\ref{fig:symmZs}.

Unlike in f(R) theories, the non-linear terms in the Klein-Gordon equation for Symmetron models lead to anti-screening effects (negative regions of the plots, c.f. Figs. \ref{fig:symmassb033} and   \ref{fig:symmZs}), or more precisely, there are configurations of interacting wave modes
that instead of driving the theory towards GR, they drive the theory away from it. In fact, from Eq.~(\ref{D2dI}), all triangular configurations with
\begin{equation}\label{AScond}
k_1^2 + k_2^2 < k_{AS}^2(a),     
\end{equation}
where the anti-screening wavenumber is defined as
\begin{equation}\label{ASscaleDef}
k_{AS}^2(a) \equiv  \frac{\kappa_3 \beta a^2}{\beta_2} - 2 m^2 a^2,     
\end{equation}
will contribute with a negative source to the 
second order screening as long as the RHS of the above equation is positive. 
We notice that for the Symmetron model, the anti-screening effects appear at scales below $k_{M_1}$, although there is not an {\it a priori} 
evident reason for this to happen and it might be the case that other Einstein frame posed theories show anti-screening effects above $k_{M_1}$ as well.

In Fig.~\ref{fig:SymmD2compare} we show an equivalent plot to Fig.~\ref{fig:F4D2compare}, where in the upper panel we show the growth functions $\bar{D}^{(2)}$ and 
$\bar{D}^{(2)}_\text{S}$, and in the lower panel their ratio. Regions where the ratios are greater than 1 correspond to wave number configurations that act as anti-screening, that is that enhance the MG fifth force.

\begin{figure}
	\begin{center}
	\includegraphics[width=3.2 in]{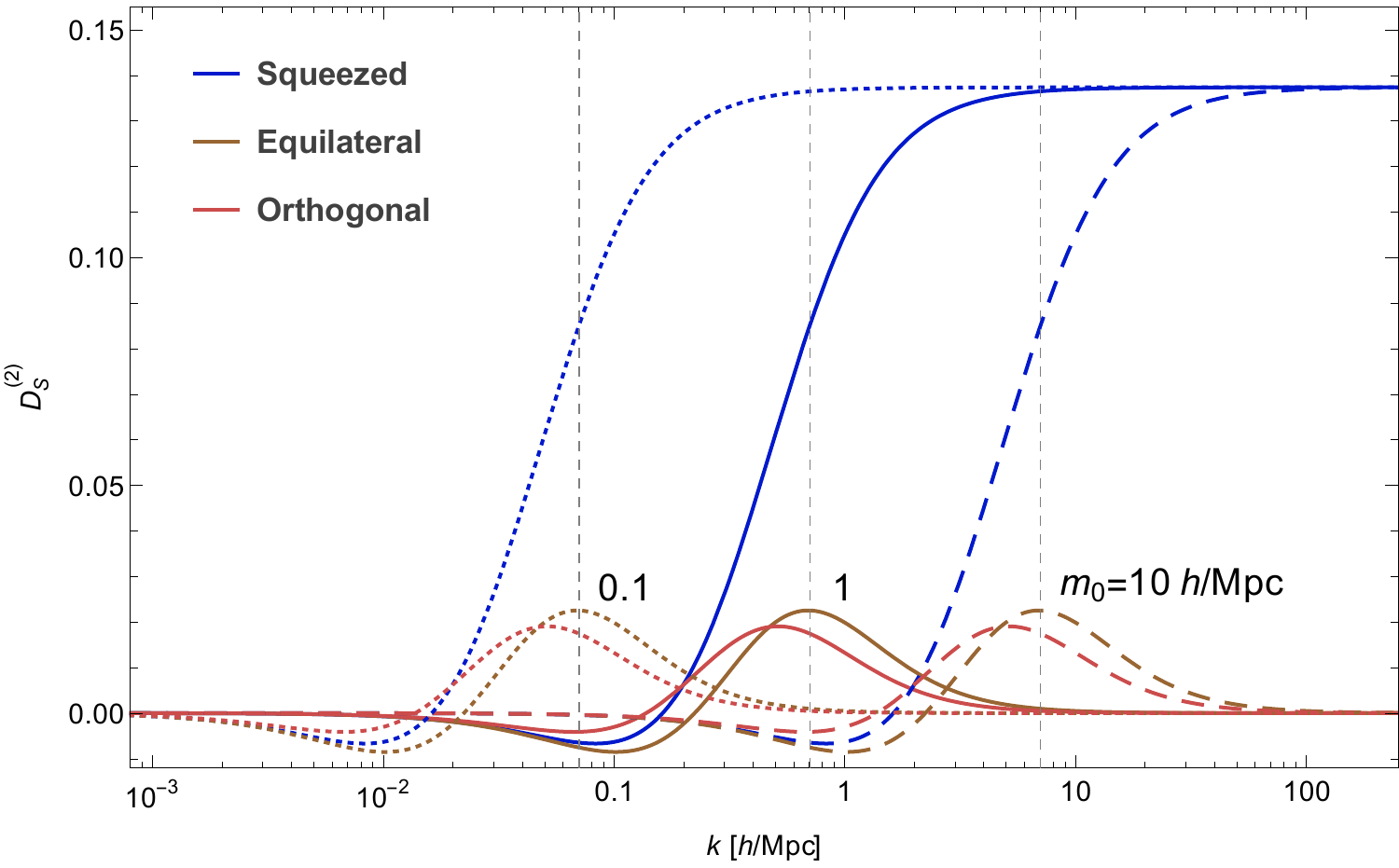}
	\includegraphics[width=3.2 in]{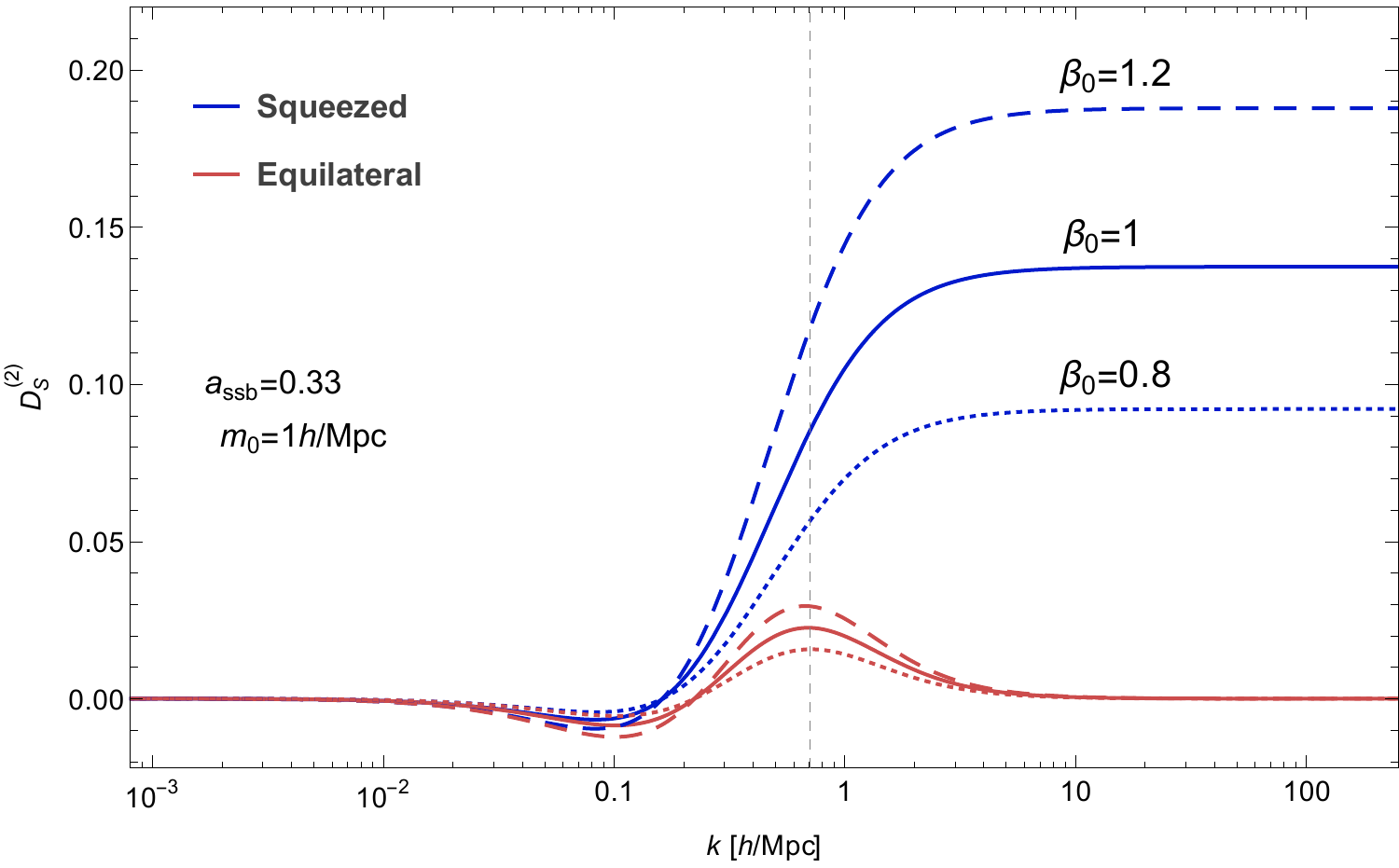}
	\caption{$ \bar{D}^{(2)}_\text{S}$ functions for the Symmetron models with  $a_{ssb}=0.33$, and considering different triangle configurations. The vertical lines are located at $k=a \sqrt{M_1}$, showing
	the characteristic scale at which the screenings are present. The upper
	panel shows the model with $\beta_0=1$ and different values of $m_0$, while the bottom panel shows the specific model with $m_0=1\, h/\text{Mpc}$ and different values of $\beta_0$.
	\label{fig:symmassb033}}
	\end{center}
\end{figure}

\begin{figure}
	\begin{center}
	\includegraphics[width=3.4 in]{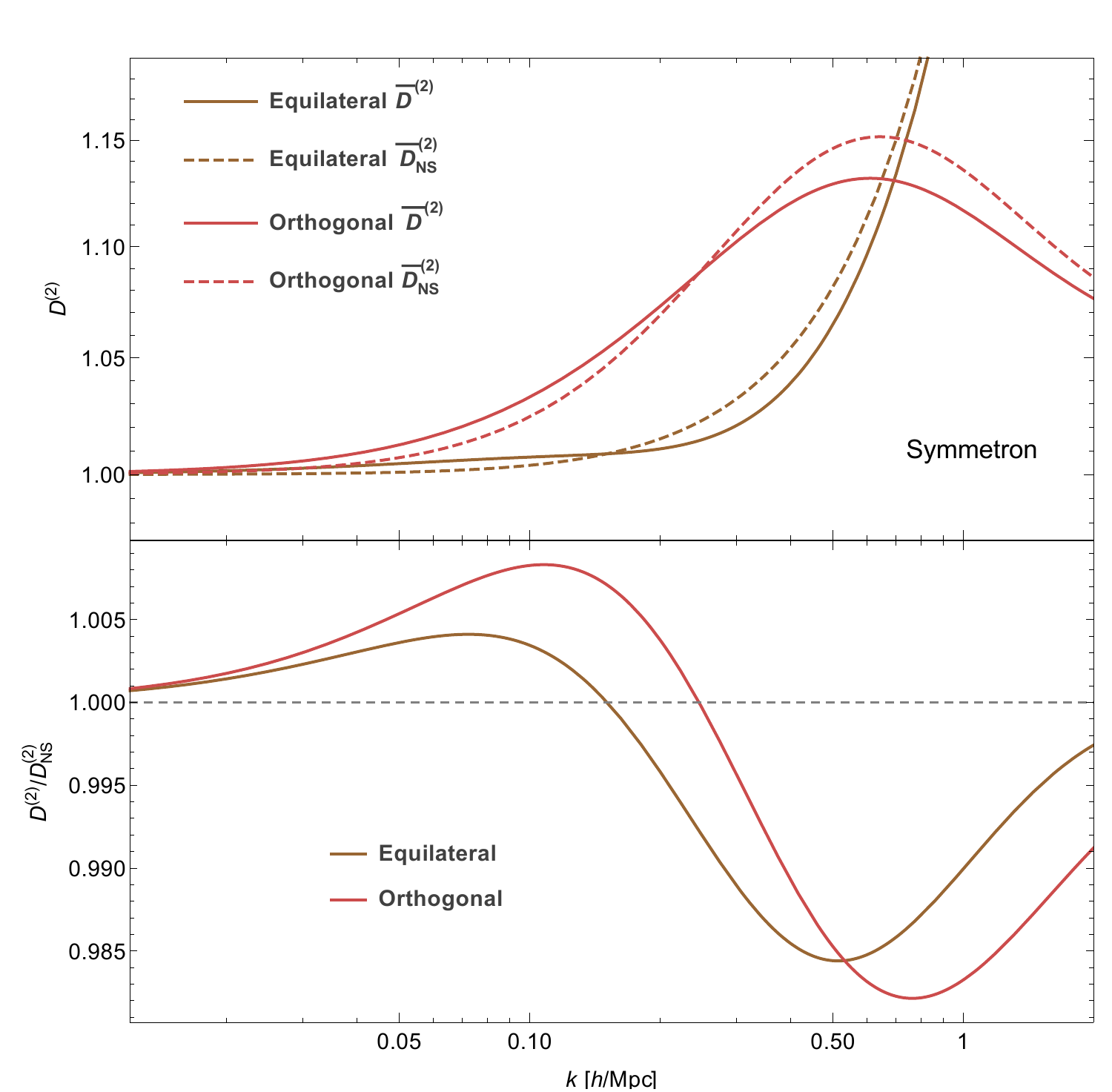}
	\caption{Normalized second order growth functions $ \bar{D}^{(2)}$ (solid curves) and $ \bar{D}^{(2)}_{NS}$ (dashed curves) for symmetron model with  $a_{ssb}=0.33$, $m_0=1 h/\text{Mpc}$ and $\beta=1$
	at redshift $z=0$. We consider equilateral (brown) and orthogonal (red) triangle configurations. 
	The bottom panel shows the ratio among them. Regions with values greater than 1, correspond to wave numbers for which the growth is enhanced by the screening contributions.
	\label{fig:SymmD2compare}}
	\end{center}
\end{figure}

\begin{figure}
	\begin{center}
	\includegraphics[width=3.2 in]{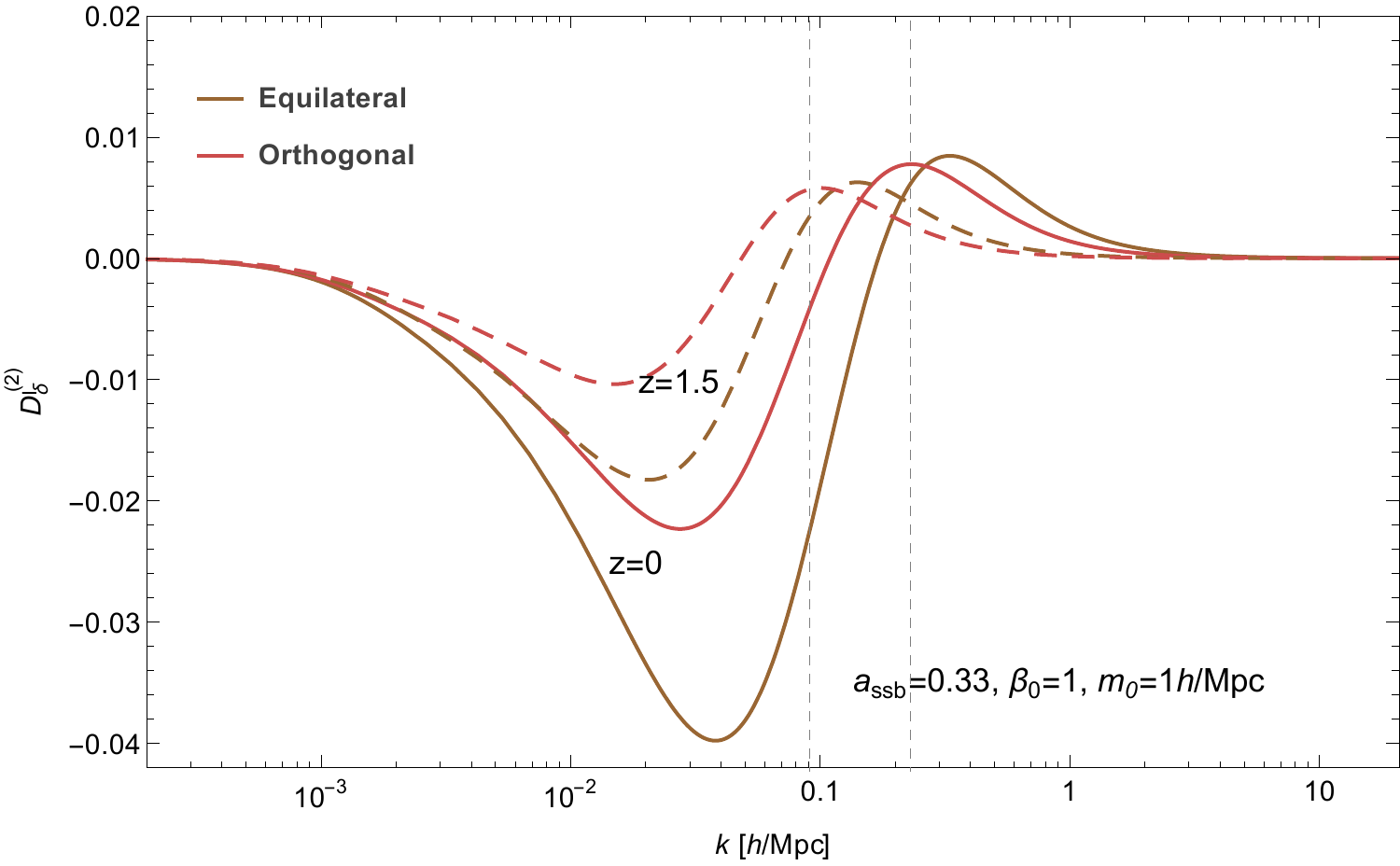}
	\caption{$ \bar{D}^{(2)}_\text{S}$ functions for the Symmetron models with  $a_{ssb}=0.33$, $m_0=1 h/\text{Mpc}$,  $\beta_0=1$,
	$\hat{n}=0.25$ and $\hat{m} = 0.5$, and considering equilateral and orthogonal triangle configurations. The vertical lines are located at $k=a \sqrt{M_1}$, showing
	the characteristic scale at which the screenings are present. 
	\label{fig:symmZs}}
	\end{center}
\end{figure}

The situation is similar when considering third order screening growth functions $D^{(3)}_{S}(\vk,-\vp,\vp)$, as it is shown in Fig.~\ref{fig:D3dI}. We again note that while f(R) models always lead to positive screenings, Symmetron models poses configurations
that enhance the MG fifth force.

\begin{figure}
	\begin{center}
	\includegraphics[width=3.2 in]{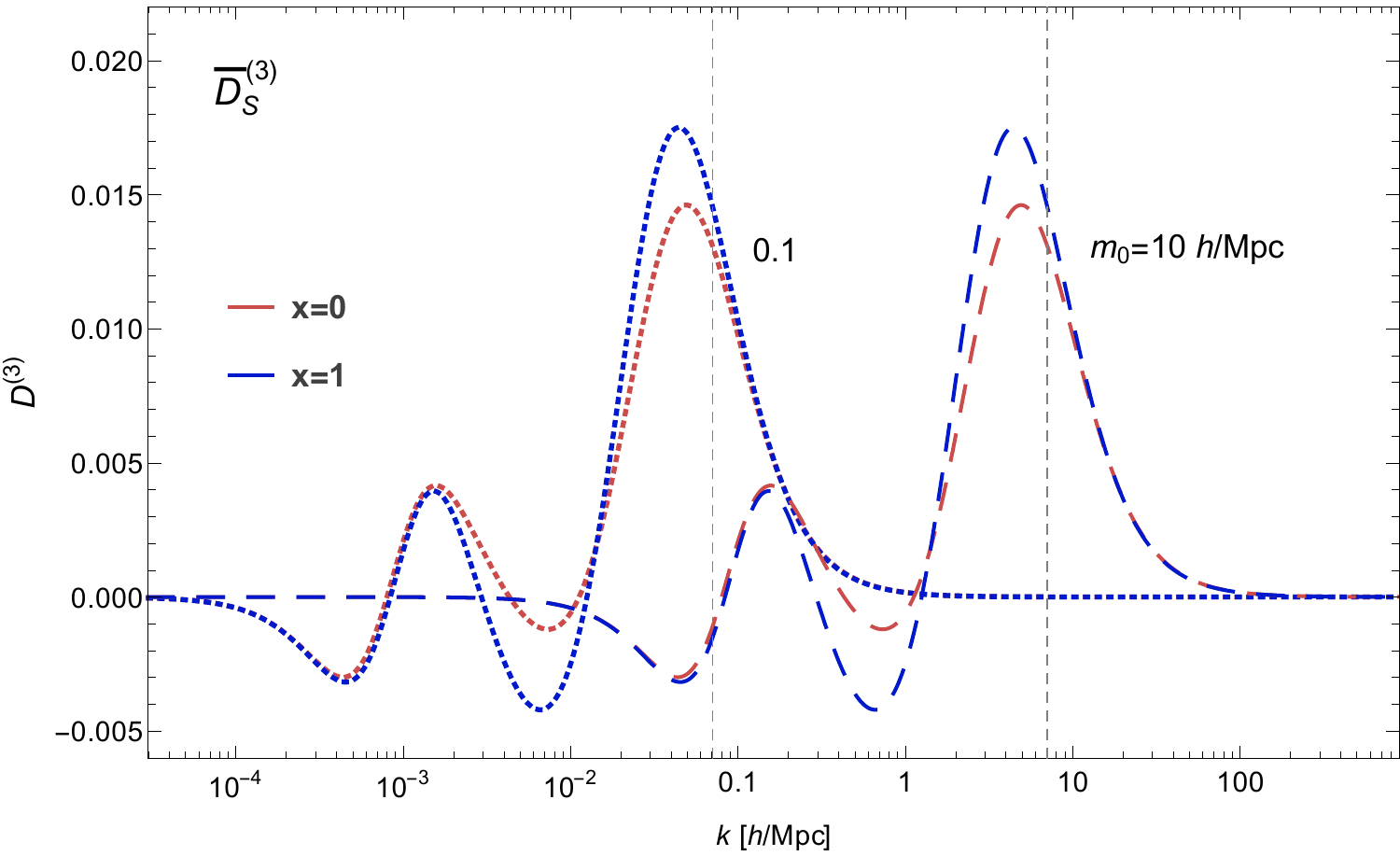}
	\caption{Third order growth function	 
	$ \bar{D}^{(3)}_\text{S}(\vk,-\vp,\vp)$ for Symmetron models with $\beta=1$ and $a_{ssb}=0.33$ and two different masses $m_0=0.1,\,10 h/\text{Mpc}$. The vertical lines are located at $k_{M_1}$ for each model, showing the characteristic scale at which the screenings are present.
	\label{fig:D3dI}}
	\end{center}
\end{figure}

We may think of the case of the (original) chameleons (\cite{Khoury:2003aq}) defined in the Einstein frame. 
Here we have a conformal factor $C(\varphi) = e^{\varphi/M} \simeq 1 + \varphi/M$ and a potential $V(\varphi)$ 
decaying with the scalar field. The linear conformal coupling implies that functions $\beta_n$ vanish for $n \geq 2$ (or at least, if we consider the full coupling $e^{\varphi/M}$, they are quite small compared to $\beta$) 
and by virtue of Eqs.~(\ref{M2}) and (\ref{M3}), the functions $M_2$ and $M_3$ become scale independent. 
Moreover, the effective mass in these models decays with time, 
implying that function $\kappa_3(a)$ is negative, while function $\kappa_4(a)$ positive. 
Then $M_2$ and $M_3$ are always positive. Hence, the growth functions will have the same kind of
 behaviour than that we observe in f(R). Equivalently, this can be observed at second order from the anti-screening wavenumber in Eq.~(\ref{ASscaleDef}), that turns out to be always negative, and hence no scale contributes to  anti-screening configurations. Contrary to this, in Symmetrons the effective mass 
grows with time and therefore $\kappa_3$ is positive. But $\beta_2$ is different 
from zero because the conformal coupling is  quadratic in the scalar field. 
Henceforth, we observe this mix of screening and anti screenings effects in PT.

\subsection{Vainshtein screening - the DGP case}

The source $S^{(2)}_\text{S}$ in Eq.~(\ref{D2dI}) for the DGP and cubic Galileons depends only on the angle $x=\hat{\vk}\cdot\hat{\vp}$ of the triangle configuration, and not on the scales $k$ and $p$. This is because the mass of the scalar field is zero in these cases. Indeed, combining Eqs.~(\ref{D2dI}) and (\ref{M2DGP}) we obtain
\begin{equation}
\mathcal{S}^{(2)}_\text{S} = \frac{Z_1(a)D_+^2(a)}{\beta^2(a)} (1-x^2),
\end{equation}
where we notice that the linear growth $D_+(a)$ is also scale independent.

In Fig.~\ref{fig:D2and3DGP} we show  $\bar{D}^{(2)}_S$  for the normal branch DGP with a crossover scale equals the Hubble size, $r_c=H_0^{-1}$. 
For both cubic Galileons and DGP, it is a function depending only on $x$ and time.

On the other hand, the symmetrized $M_3$ function, again in the double squeezed configuration, becomes
\begin{equation}\label{M3DGP_2}
M_3(\vk,-\vp,\vp) = \frac{Z_1(a)}{2 a^2 \beta^4(a) A_0}[k^2 p^4 (1-x^2) - k^3 p^3 x(1-x^2)].
\end{equation}
Because the second term on the RHS of the above equation is odd in $x$, it does not contribute when $D^{(3)}$ is integrated in Eq.~(\ref{R1f}) to obtain loop statistics. Hence we do not consider it here.
The third order screening growth $\bar{D}^{(3)}_S$ turn out to depend on the scale, but only through the ratio $p/k$. In Fig.~\ref{fig:D2and3DGP} we show this dependence for $p=k$, and the limiting cases $p\gg k$ and $p\ll k$. 

We observe that the second and third order screening growths always act to attenuate the fifth force that modifies Newtonian gravity. 

Recently, the authors of \citep{Ogawa:2018srw} found matter configurations that yield anti-screening responses in the cubic Galileon model.  In PT we found that these are not present, at least up to third order in matter fluctuations. 

\begin{figure}
	\begin{center}
	\includegraphics[width=3.2 in]{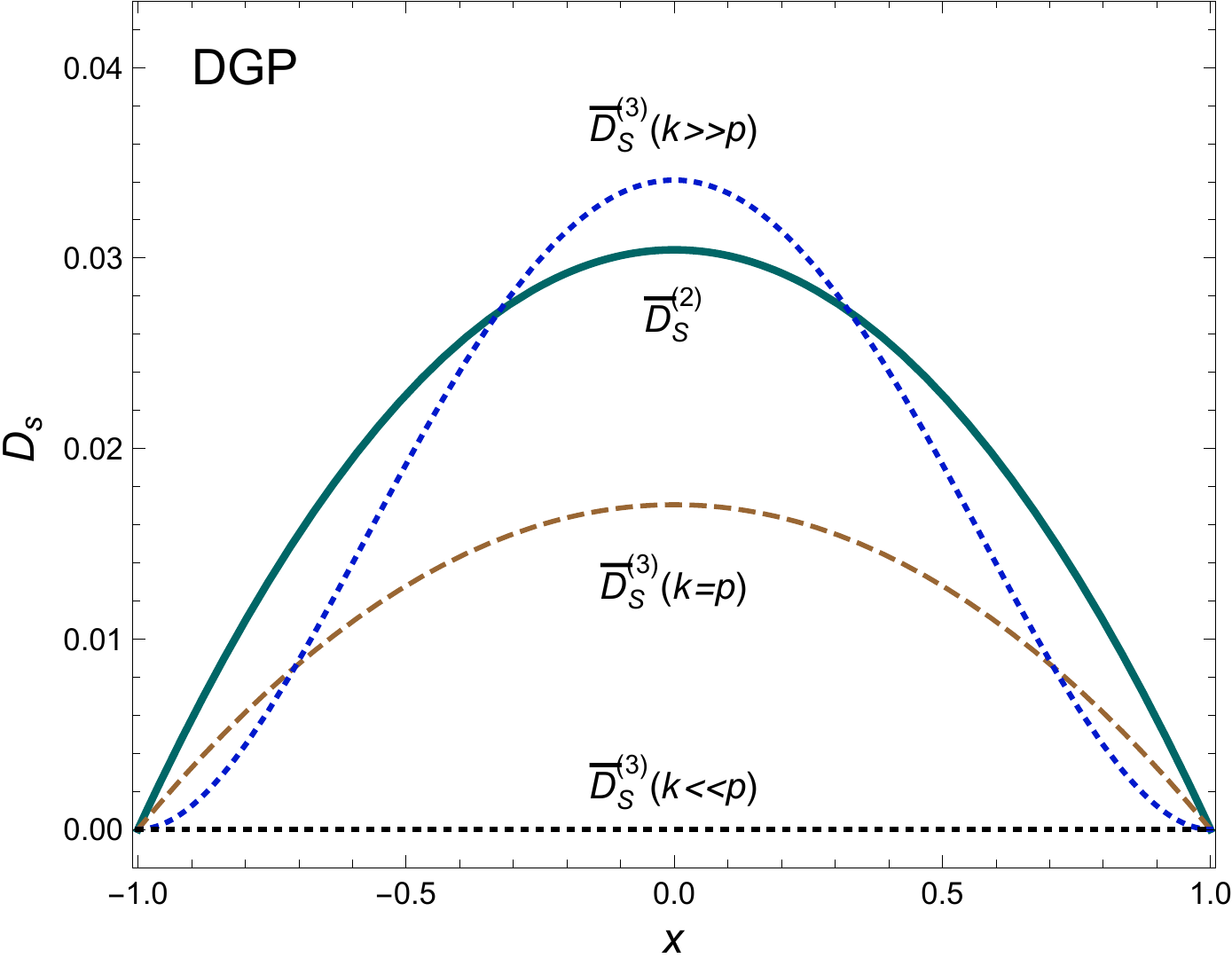}
	\caption{ $ \bar{D}^{(2)}_\text{S}$ and 
	$ \bar{D}^{(3)}_\text{S}$ functions  as a function of cosine angle $x=\hat{\vp}\cdot \hat{\vk}$ for DGP model with 
	crossover scale $r_c=1/H_0$. We state that the behaviour is qualitatively the same for cubic Galileons. 
	\label{fig:D2and3DGP}}
	\end{center}
\end{figure}

\end{section}
  
\section{Conclusions}\label{sec:conclusions}

In this article we present a formalism to study screening mechanisms in modified theories of gravity via perturbative methods. 
We use a redefinition of the scalar degree of freedom that permit us to recast the Einstein frame perturbation equations to the Jordan frame, for which we have at hand a previously developed  theory for matter clustering in MG, that we are then able to apply. In spite of the fact that screening mechanisms are nonlinear phenomena, our perturbative approach give us an analytical tool to probe and understand features in screening mechanisms. This allow us to compare several theoretical models and  to identify features which can be used to differentiate among them through their screening mechanisms.

An interesting result we obtain is that the perturbative screenings do not always act in the “screening direction”; that is, although the Symmetron power spectrum, when considering only the source $\mathcal{S}_1$ in Eq.~(\ref{D3symm}), has less screening contributions than the full loop curves, they are actually closer to the GR power spectrum, as shown in Fig.~\ref{fig:symmetronPS}. The reason for this behaviour has its roots in the non-linearities of the Klein-Gordon equation in which the couplings provide anti-screening effects, among other effects.

 We identify the emergence of a natural screening wavenumber, $k_{M_1} = a \sqrt{M_1(a)}$, for {\it weak coupling} and {\it large mass} screening models that serve us to identify a scale of appearance of the screenings effects, as shown in Figs.~(\ref{fig:F4F8F12}-\ref{fig:symmZs}). Although this represents only an approximate number, it could be useful for a rapid identification of screening occurrence.

Our computations show that in f(R) theories the  nonlinearities  of the  Klein-Gordon  equation  lead  to  screening  for  any   configuration. This is because the screening growing functions always take positive values. Unlike f(R) theories, the non-linear terms in the Klein-Gordon equation for Symmetron models lead to anti-screening effects. That is, there are configurations of interacting wave modes that instead of driving the theory towards General Relativity, they drive the theory away from it.
We trace back this behaviour to both the quadratic conformal coupling and the effective mass of the theory that grows with time. In contrast, for the standard chameleon defined in the Einstein frame, the conformal factor is linear in the scalar field and its mass decays with time; as an outcome, the screening features of this model are qualitatively the same as those in f(R) theories. 
On the other hand, for  the  DGP  and  cubic Galileon models we find no signatures of anti-screening up to third order in perturbation theory. Moreover, their second order growth  functions become trivial as they do not depend on the size of the triangle configurations, but only on one of the angles that define them. That is, they become scale independent, that we notice is a consequence of the vanishing mass in these models.

Screenings mechanisms leave imprints in the quasi-nonlinear matter power spectra, that may represent a way to distinguish different gravity 
theories that otherwise behave in a very similar way at background and linear cosmological levels. Our present study sheds light towards finding 
smoking guns among the different screening models within MG theories. This is especially important to validate MG N-body simulations with theory to 
later compare with forthcoming precision data from large scale galaxy surveys as eBOSS, DESI, EUCLID, and LSST.

\begin{acknowledgements}
DFM thanks the Research Council of Norway for their support, and the resources provided by 
UNINETT Sigma2 -- the National Infrastructure for High Performance Computing and 
Data Storage in Norway.
This paper is based upon work from the COST action CA15117 (CANTATA), supported by COST (European Cooperation in Science and Technology).  JLCC and AA acknowledge financial supported by Consejo Nacional de Ciencia y Tecnolog\'ia (CONACyT), Mexico, under grants: 283151 and Fronteras Project 281. 
\end{acknowledgements}

\appendix

\begin{section}{Third order growth functions}\label{app:3rdOrder}

In this appendix we display the sources of the third order growth function $D^{(3)s}(\vk,-\vp,\vk)$ differential 
equation of Eq.~(\ref{D3symm}). First we define the normalized second order growth of matter fluctuations
\begin{equation}
D^{(2)}(\vk_1,\vk_2) = D^{(2)}_\text{NS}(\vk_1,\vk_2) - D^{(2)}_\text{S}(\vk_1,\vk_2) 
\end{equation}
with $D^{(2)}_\text{NS}$ and $D^{(2)}_\text{S}$ solutions to Eqs.~(\ref{D2NoSc}) and (\ref{D2dI}). We also will use 
the growth function for scalar field $\chi$.
\begin{align}
D^{(2)}_\chi(\vk_1,\vk_2) &=  D^{(2)}(\vk_1,\vk_2) + (1+(\hat{\vk}_1 \cdot \hat{\vk}_2)^2) D_+(\vk_1) D_+(\vk_1) \nonumber\\
 &- \frac{2A_0}{3} \frac{M_2^\text{FL}(\vk_1,\vk_2)}{3 \Pi(k_1) \Pi(k_2)},
\end{align}
with the frame-lagged $M^\text{FL}_2$ function 
\begin{equation}\label{M2FL}
M^\text{FL}_2(\vk_1,\vk_2) = M_2(\vk_1,\vk_2) + \frac{9C}{4 \beta^2 A_0^2}\mathcal{K}_\text{FL}(\vk_1,\vk_2)\Pi(k_1)\Pi(k_2). 
\end{equation}
The sources of Eq.~(\ref{D3symm}) are given by
\begin{align}
 \mathcal{S}_1 &=  D_+(p)\left(A(p) + \T - A(k)\right) D^{(2)}(\vp,\vk) 
                \left( 1- \frac{(\vp \cdot (\vk + \vp))^2}{p^2 |\vp+\vk|^2}\right)  \nonumber \\* & + (\vp \rightarrow -\vp), \\
 \mathcal{S}_2 &= - D_+(p)\left(A(p) + A(|\vk + \vp|) - 2A(k)\right) D^{(2)}(\vp,\vk)  
                \nonumber \\* & + \left(2A(k) -  A(p) - A(|\vk + \vp|) \right) D_+(k) D_+^2(p) \frac{(\vk \cdot \vp)^2}{k^2p^2}  \nonumber\\*
               &- \big( A(|\vk + \vp|) -  A(k) \big) D_+(k) D_+^2(p) 
               -  \Bigg[ \frac{M_1(\vk + \vp)}{3\Pi(|\vk + \vp|)}\mathcal{K}^{(2)}_\text{FL}(\vp,\vk) \nonumber\\*
&                -  \left(\frac{2 A_0}{3}\right)^2  \frac{M_2(\vp,\vk) |\vk + \vp|^2/a^2}{6\Pi(|\vk + \vp|)\Pi(k)\Pi(p)} \Bigg]  D_+(k) D_+^2(p)  \nonumber\\*
& + \frac{M_1(k)}{3 \Pi(k)} \Bigg[ \left( 2\frac{(\vp \cdot (\vk + \vp))^2}{p^2 |\vp+\vk|^2}  - \frac{\vp\cdot(\vk+\vp)}{p^2}\right)\nonumber\\*
&     (A(p) - A_0) D^{(2)}(\vp,\vk) D_+(p) +\left( 2 \frac{(\vp \cdot (\vk + \vp))^2}{p^2 |\vp+\vk|^2}  - \frac{\vp\cdot(\vk+\vp)}{|\vk+\vp|^2}\right) \nonumber\\*
&   \times (A(|\vk+\vp|) - A_0) D^{(2)}_\chi(\vp,\vk) D_+(p) +   \nonumber\\*
  &  3 \frac{(\vk \cdot \vp)^2}{k^2 p^2} 
  \big( A(k) +  A(p) - 2 A_0 \big) D_+(k) D_+^2(p)   \Bigg]    + (\vp \rightarrow -\vp),         \\
\mathcal{S}_3 &= -\frac{k^2/a^2}{6\Pi(k)} \mathcal{K}_{\delta I}^{(3)s} (\vk,-\vp,\vp) D_+(k) D_+^2(p),     
\end{align}
with the third order kernel of screenings
\begin{align}
& \mathcal{K}_{\delta I}^{(3)s} (\vk,-\vp,\vp) = 2\left(\frac{2A_0}{3}\right)^2 \frac{M_2(\vk,0)}{\Pi(k)\Pi(0)} 
\nonumber\\
&  + \frac{(2A_0/3)^3}{3\Pi^2(p)\Pi(k)} \Bigg[M_3(\vk,-\vp,\vp) - \frac{M_2(\vk,0)M_2^\text{FL}(-\vp,\vp)}{\Pi(0)} \Bigg] \nonumber\\
& +\left(\frac{2A_0}{3}\right)^2 \frac{M_2(-\vp,\vk+\vp)}{\Pi(k)\Pi(|\vk+\vp|)} \Bigg(1+(\hat{\vk}\cdot\hat{\vp})^2 + 
\frac{D^{(2)}(\vp,\vk)}{D_+(p)D_+(k)} \Bigg)\nonumber\\
& + \frac{(2A_0/3)^3}{3\Pi^2(p)\Pi(k)} 
\Bigg[M_3(\vk,-\vp,\vp) - \frac{M_2(-\vp,\vk+\vp)M_2^\text{FL}(\vk,\vp)}{\Pi(|\vk+\vp|)} \Bigg]\nonumber\\
& +\left(\frac{2A_0}{3}\right)^2 \frac{M_2(\vp,\vk-\vp)}{\Pi(k)\Pi(|\vk-\vp|)} \Bigg(1+(\hat{\vk}\cdot\hat{\vp})^2 + 
\frac{D^{(2)}(-\vp,\vk)}{D_+(p)D_+(k)} \Bigg)\nonumber\\
& + \frac{(2A_0/3)^3}{3\Pi^2(p)\Pi(k)} 
\Bigg[M_3(\vk,-\vp,\vp) - \frac{M_2(\vp,\vk-\vp)M_2^\text{FL}(\vk,-\vp)}{\Pi(|\vk-\vp|)} \Bigg].
\end{align}
We should note that these functions are valid for symmetrized $M_2$ and $M_3$ functions.

\end{section}

\bibliography{refs}

\end{document}